  \documentclass[longauth]{aa} 
 
\usepackage{graphicx}
 \usepackage[]{natbib}
\usepackage{txfonts}
\usepackage{color}
\definecolor{ladrillo}{RGB}{102, 0.0, 0.}
\usepackage{url}
\bibpunct{(}{)}{;}{a}{}{,} 
\begin{document}

   \title{The Gaia-ESO Survey: metallicity and kinematic trends in the Milky Way bulge \thanks{Based on observations made with the ESO/VLT, at Paranal Observatory, under program 188.B-3002, The Gaia-ESO Public Spectroscopic Survey}}

\author{A.~Rojas-Arriagada \inst{\ref{inst1}}
  \and A.~Recio-Blanco \inst{\ref{inst1}}
  \and V.~Hill \inst{\ref{inst1}}
  \and P.~de Laverny \inst{\ref{inst1}}
  \and M.~Schultheis \inst{\ref{inst1}}
  \and C.~Babusiaux \inst{\ref{inst19}}
  \and M.~Zoccali\inst{\ref{inst5},\ref{inst6}} 
  \and D.~Minniti\inst{\ref{inst5},\ref{inst6},\ref{inst7},\ref{inst21}}
  \and O.~A. Gonzalez\inst{\ref{inst8}}
  \and S.~Feltzing \inst{\ref{inst15}}
  \and G.~Gilmore \inst{\ref{inst3}}
  \and S.~Randich \inst{\ref{inst4}}  
  \and A.~Vallenari  \inst{\ref{inst18}}
  \and E.~J. Alfaro \inst{\ref{inst11}}  
  \and T.~Bensby \inst{\ref{inst15}}
  \and A.~Bragaglia \inst{\ref{inst10}}
  \and E.~Flaccomio \inst{\ref{inst9}}
  \and A.~C. Lanzafame \inst{\ref{inst13}}
  \and E.~Pancino \inst{\ref{inst10}}
  \and R.~Smiljanic \inst{\ref{inst16},\ref{inst17}}
  \and M.~Bergemann  \inst{\ref{inst3}}
  \and M.~T. Costado  \inst{\ref{inst11}}
  \and F.~Damiani \inst{\ref{inst9}}
  \and A.~Hourihane \inst{\ref{inst3}}
  \and P.~Jofr\'e  \inst{\ref{inst3}}
  \and C.~Lardo  \inst{\ref{inst10}}
  \and L.~Magrini \inst{\ref{inst4}}
  \and E.~Maiorca \inst{\ref{inst4}}
  \and L.~Morbidelli \inst{\ref{inst4}}
  \and L.~Sbordone  \inst{\ref{inst12}}
  \and C.~C.~Worley \inst{\ref{inst3}}
  \and S.~Zaggia \inst{\ref{inst18}}
  \and R.~Wyse \inst{\ref{inst20}}
}
\institute{
Laboratoire Lagrange (UMR 7293), Universit\'e  Nice Sophia Antipolis, CNRS, Observatoire de la C\^ote d'Azur, CS 34229, 06304 Nice, cedex 04, France  \email{arojas@oca.eu} \label{inst1}
\and
GEPI, Observatoire de Paris, CNRS, Universit\'e Paris Diderot, 5 Place Jules Janssen, 92190 Meudon, France \label{inst19}
\and
Instituto de Astrof\'{i}sica, Facultad de F\'{i}sica, Pontificia Universidad Cat\'olica de Chile, Av. Vicu\~na Mackenna 4860, Santiago, Chile \label{inst5}
\and
The Milky Way Millennium Nucleus, Av. Vicu\~{n}a Mackenna 4860, 782-0436 Macul, Santiago, Chile\label{inst6}
\and
Vatican Observatory, V00120 Vatican City State, Italy \label{inst7}
\and 
Departamento de Ciencias Fisicas, Universidad Andres Bello, Republica 220, Santiago, Chile\label{inst21}
\and
European Southern Observatory, A. de Cordova 3107, Casilla 19001, Santiago 19, Chile  \label{inst8}
\and
Lund Observatory, Department of Astronomy and Theoretical Physics, Box 43, SE-221 00 Lund, Sweden \label{inst15}
\and
Institute of Astronomy, University of Cambridge, Madingley Road, Cambridge CB3 0HA, United Kingdom \label{inst3}
\and
INAF - Osservatorio Astrofisico di Arcetri, Largo E. Fermi 5, 50125, Florence, Italy \label{inst4}
\and
INAF - Osservatorio Astronomico di Padova, Vicolo dell'Osservatorio 5, 35122 Padova, Italy \label{inst18}
\and
Instituto de Astrof\'{i}sica de Andaluc\'{i}a-CSIC, Apdo. 3004, 18080, Granada, Spain \label{inst11}
\and
INAF - Osservatorio Astronomico di Bologna, via Ranzani 1, 40127, Bologna, Italy \label{inst10}
\and
INAF - Osservatorio Astronomico di Palermo, Piazza del Parlamento 1, 90134, Palermo, Italy \label{inst9}
\and
Dipartimento di Fisica e Astronomia, Sezione Astrofisica, Universit\'a di Catania, via S. Sofia 78, 95123, Catania, Italy \label{inst13}
\and
Department for Astrophysics, Nicolaus Copernicus Astronomical Center, ul. Rabia\'{n}ska 8, 87-100 Toru\'{n}, Poland \label{inst16}
\and 
European Southern Observatory, Karl-Schwarzschild-Str. 2, 85748 Garching bei M\"unchen, Germany \label{inst17}
\and
ZAH - Landessternwarte Heidelberg, Königstuhl 12, D-69117, Heidelberg, Germany \label{inst12}
\and
Department of Physics and Astronomy, Johns Hopkins University, 3400 North Charles Street, Baltimore, MD 21218 \label{inst20}
 }

   \date{Received...; accepted...}

   \newcommand{\teff}{T$_{\rm eff}~$}
   \newcommand{\logg}{$\log{g}~$}
   \newcommand{\feh}{$\rm [Fe/H]~$}
   \newcommand{\met}{${\rm [M/H]}~$}
   \newcommand{\aabun}{${\rm [\alpha/Fe]}$}
   \newcommand{\kms}{km~s$^{-1}$}
   \newcommand{\vrad}{${\rm V_{rad}}$}

 
  \abstract
   {}
   {Observational studies of the Milky Way bulge are providing increasing evidence of its complex chemo-dynamical patterns and morphology.  Our intent is to use the iDR1 Gaia-ESO survey data set to provide new constraints on the  metallicity and kinematic trends of the Galactic bulge, exploring the viability of the currently proposed formation scenarios.}
   {We analyzed the stellar parameters and radial velocities of $\sim1200$ stars in five bulge fields wich are located in the region $-10^{\circ}<l<7^{\circ}$ and $-10^{\circ}<b<-4^{\circ}$. 
We use VISTA Variables in The Via Lactea (VVV) photometry to verify the internal consistency of the atmospheric parameters recommended by the Gaia-ESO Survey consortium. As a by-product, we obtained reddening values using a semi-empirical T$_{\rm eff}$-color calibration. We constructed the metallicity distribution functions and combined them with photometric and radial velocity data to analyze the properties of the stellar populations in the observed fields.}
   {From a Gaussian decomposition of the metallicity distribution functions, we unveil a clear bimodality in all fields, with the relative size of components depending of the specific position on the sky. In agreement with some previous studies, we find a mild gradient along the minor axis (-0.05 dex/deg between $b=-6^{\circ}$ and $b=-10^{\circ}$) that arises from the varying proportion of metal-rich and metal-poor components. The number of metal-rich stars fades in favor of the metal-poor stars with increasing $b$. The K-magnitude distribution of the metal-rich population splits into two peaks for two of the analyzed fields that intersects the near and far branches of the X-shaped bulge structure. In addition, two lateral fields at $(l,b)=(7,-9)$ and $(l,b)=(-10,-8)$ present contrasting characteristics. In the former, the metallicity distribution is dominated by metal-rich stars, while in the latter it presents a mix of a metal-poor population and and a metal-intermediate one, of nearly equal sizes. Finally, we find systematic differences in the velocity dispersion between the metal-rich and the metal-poor components of each field.
   }
   {The Gaia-ESO Survey iDR1 bulge data show chemo-dynamical distributions that are consistent with varying proportions of stars belonging to (i) a metal-rich boxy/peanut X-shaped component, with bar-like kinematics, and (ii) a metal-poor more extended rotating structure with a higher velocity dispersion that dominates far from the Galactic plane. These first GES data already allow  studying the detailed spatial dependence of the Galactic bulge populations, thanks to the analysis of individual fields with relatively high statistics.}

   \keywords{Galaxy: formation, abundances, bulge, stellar content -- stars: abundances
               }
   \maketitle
%

\section{Introduction}
\label{sec:introduccion}

The bulge is a major component of our Galaxy, comprising approximately a quarter of its total stellar mass \mbox{($\textmd{M}_{bulge}\sim1.8\times10^{10}\textmd{M}_{\odot}$, \citeauthor{sofue_masa_bulbo} \citeyear{sofue_masa_bulbo})}. Bulges are very common structures in galaxies, which in cosmological scales harbor approximately 70\% of the stellar mass \citep{fukugita}. Because they consist of most of the oldest stars in a galaxy, studying bulges represents a valuable opportunity of understanding the chemo-dynamical processes involved in the general formation of the host galaxy. Furthermore, because of its proximity, the Galactic bulge presents an ideal opportunity of performing detailed observations for hypothesis-testing on a star-by-star basis, an advantage that is not available for external bulges, except for M31.\\
In the past years, Milky Way bulge studies have been the object of intense debate,  when it became evident that it was impossible to consider it as a single stellar population \citep{babusiaux2010,hill2011,bensby2011,argos_3}. Its complex internal structure and kinematic and chemical patterns have inspired several studies in the community during the past decade that aimed at explaining its formation and evolution. The three main formation scenarios currently invoked are (i) in situ formation via dissipative collapse of a protogalactic gas cloud in a free-fall time scale \citep{eggen_1962}, (ii) accretion of substructures, disk clumps or external building blocks in a $\Lambda$CDM context \citep{immeli_2004,scannapieco_2003}, and (iii) secular formation from disk material through bar formation, vertical instability, buckling,  and fattening, which produces a structure commonly called pseudobulge \citep{pfenniger_1990,kormendy_kennicutt_review}.\\
The increasing amount of observational evidence supports different scenarios in an apparently contradictory way.\\
The shape and kinematics of the Milky Way bulge both support the existence of a bar. The bulge presents a boxy/peanut appearance as displayed in 2MASS star counts \citep{lopez_corredoira_boxy} and COBE-DIRBE near-infrared light distribution \citep{dwek}. Star counts using red clump (RC) stars characterize the bar as a triaxial structure of $\sim3.5$ kpc in length that points at the first quadrant at $\sim25^{\circ}$  with respect to the Sun-Galactic center line of sight \citep{rattenbury_barra}. Several models have subsequently interpreted this boxy structure as an edge-on buckled bar \citep{zhao_barra,fux_barra,shen2010}. 
Boxy bulges present a characteristic cylindrical rotation, namely, a nearly constant stellar rotation speed with height above the Galactic plane. Classical bulges rotate more slowly at increasing height \citep{falcon_barroso}. After studies of various probes (K giants, globular clusters, PNe), the first and recent large data set of systematically collected radial velocities in the Galactic bulge was assembled by the BRAVA project \citep{Rich_brava_I}, and showed indistinguishable rotation curves in strips at $b=-4^\circ$, $b=-6^\circ$, and $b=-8^\circ$ ($l\pm10$). BRAVA velocities are fitted nicely by the model of \citet{shen2010}, which is able to reproduce a boxy/peanut bulge just from accreted stars from the disk and constrains any possible classical component to be less than 8\% of the disk mass. This could challenge some attempts to detect it, even in large data samples.\\

Furthermore, the Galactic bulge has been shown to display an X-shape. In an extensive analysis using star counts of RC members in a large area, \citet{mcwilliam_zoccali2010} using 2MASS, and \citet{nataf2010} using OGLE-III, have revealed the existence of a bimodality in the RC magnitude at $|b|>5.5^{\circ}$. The distance between the two magnitude peaks appears to be nearly constant with longitude and decreases toward the Galactic plane. This is interpreted as evidence for an X-shaped Galactic bar. \citet{saito_x_shape} and \citet{wegg_mapas} have confirmed these findings by tracing the bulge structure in the line of sight in several slices in $l$ and $b$ using density maps. The underlying X shaped structure is seen as boxy/peanut from the Sun position because of the almost edge-on perspective. In this way, the double red clump arises from the intersection of the line of sight with the two overdensities that correspond to the X shape. In addition, as found by \citet{ness_splitted_rc}, the double red clump feature is verified only for stars with [Fe/H]$>$-0.5 dex.
Some dynamical models predict X-shaped bulges as extreme cases of boxy/peanut bulges \citep[e.g,][]{athanassoula_2005,martinez_valpuesta_2006,debattista_2006}. This type of structure has also been observed in external galaxies \citep{bureau_2006}.\\
Radial velocity measurements by \citet{rangwala_2009} in the region dominated by the bar have revealed stellar streaming motions at $(l,b)=(\pm 5,-4)$. \citet{babusiaux2014} who used RC stars in four fields with $b=0$, found streaming motions induced by the bar at $l\pm6$ that were more evident for stars with \feh$\geq-0.2$ dex. Dedicated investigations that probed radial velocity in bulge fields displaying the double RC feature have reached different conclusions. i) \citet{de_propris_2011} found no radial velocity difference between the two RC overdensities at $(l,b)=(0,-8)$, ii)  \citet{sergio_x_kine} detected two streams by comparing the bright and faint RC at $(l,b)=(0,-6)$, for which they used both radial velocities and transverse motions, and iii) \citet{uttenthaler_2012} found no statistically significant difference between the two RCs in the field at $(l,b)=(0,-10)$ in the bulge outskirts. These results compare well with predictions extracted from evolutionary N-body models \citep{athanassoula2003,ness_splitted_rc,sergio_x_kine} and highlight the complex orbital structure of the bar component.\\
Finally, a very important piece of evidence, still largely unexplored, is the true shape of the metallicity distribution function (MDF) across the whole bulge region and its relation with the different structural and kinematical observed features.
Most of the early work was mainly restricted to Baade's Window, thus preventing the detection of any variation of the MDF with ($l$,$b$). \citet {zoccali2008} presented the first complete MDF determination based on high-resolution spectra for three fields in the bulge minor axis.
Their sample revealed a metallicity gradient that was interpreted as evidence in favor of an early collapse formation scenario. Such  a gradient would be absent at latitudes lower than $b=-4^{\circ}$  \citep{rich_no_gradiente,Rich_2012_no_gradiente}. Using the same data, \citet{babusiaux2010} suggested that two  chemically and kinematically distinguishable populations might coexist in these fields. \citet{babusiaux2010} and \citet{hill2011} specifically showed that the most metal-rich, alpha-poor population in Baade's Window presents a vertex deviation that is compatible with bar-like kinematics, while the metal-poor, alpha-rich population did not, as expected for a spheroid, for example.
More recently, from a large sample of $28,000$ low to medium-resolution spectra in 28 fields, \citet{argos_3} presented a detailed MDF study that suggests there might be up to five underlying component populations. This new characterization agree with the observation reported by \citet{zoccali2008} that the metallicity gradient is produced by a change of MDF shape and not by a solid shift with latitude. It also agrees with the results of \citet{babusiaux2010} and \citet{hill2011} that suggested two chemically and kinematically distinguishable populations. These findings reveal the bulge as a very complex structure, consisting of several populations with different characteristics, origins, and possible formation mechanisms.\\
To provide new constraints on these mentioned formation scenarios and the bulge structure, we present here the first analysis of Gaia-ESO Survey (GES) bulge fields. These fields chemically and kinematically probe the stellar populations in five locations within and outside the minor axis. The structure of the paper is as follows: In Section \ref{sec:datos}, we describe the data and the selection criteria. Section \ref{sec:determinacion_reddenings} presents our new reddening determinations for the fields. The final selected sample is described in Sect. \ref{sec:seleccion_final}. Section \ref{sec: caracterizacion_vel_rad} describes the general characteristics of the radial velocity measurements, while in Sect. \ref{sec:distros_met} we present and discuss the metallicity distribution functions. The double-peaked red clump populations and their nature are in Sect. \ref{sec:rc_split}. A detailed analysis of the kinematical features of the observed populations is given in Sect. \ref{sec:kinematical_features}. Finally, Sect. \ref{sec:conclusiones} presents the discussion and our conclusions.


\section{Data}
\label{sec:datos}
The Gaia-ESO survey (GES) is a public spectroscopic survey targeting $\geq10^5$ stars that covers all the major components of the Milky Way, from halo to star-forming regions. The GES consortium \citep{GESMessenger,randich2013} works on the basis of work packages and manages a common effort in a structured way to process and analyze this large amount of data. The GES processing includes target selection, data reduction, spectrum analysis, astrophysical parameter determination, calibration, and homogenization. A detailed description of the data processing cascade and a general characterization of the data set can be found in Gilmore et al. (in prep.).\\
\begin{table} 
\centering
\caption{Characterization of the observed fields. The reddening $E(J-K)$ values are those estimated in Sect. \ref{sec:determinacion_reddenings}. The signal-to-noise ratio is derived from averaging the individual values of all stars in each field.}
\begin{tabular}{lrrrrrc}
\hline
\hline
Field name & $l$ & $b$ & SN & E(J-K) & N$_{\textmd{stars}}$ & \%Giants \\\hline
p1m4 & 1 & -4 & 90 & 0.21 & 321 & 98 \\ 
p0m6 & 0 & -6 & 82 & 0.14 & 227 & 98 \\
m1m10 & -1 & -10 & 101 & 0.06 & 227 & 74 \\
p7m9 & 7 & -9 & 81 & 0.11 & 223 & 91 \\
m10m8 & -10 & -8 & 98 & 0.06 & 222 & 79 \\ \hline
\end{tabular}
\label{tab:caract_campos}
\end{table}

\begin{figure}
\begin{center}
\includegraphics[width=9cm]{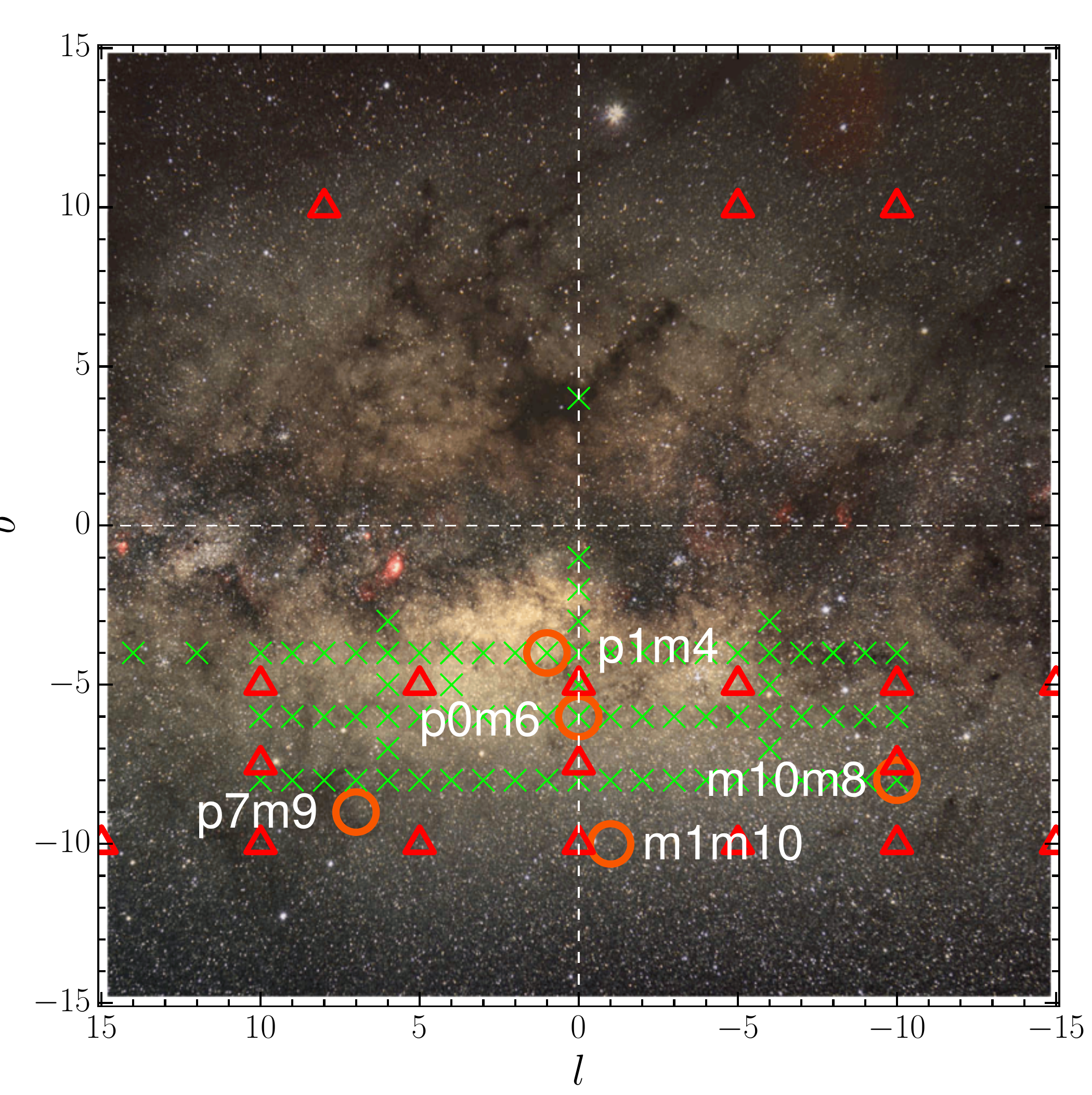}
\caption{Approximate position of the five bulge fields. To set our data in the context of recent surveys that  explored the bulge region, we also include the fields observed in the BRAVA (green crosses) and in the ARGOS survey (red triangles). All fields are overplotted on an optical image of the Milky Way bulge (\textcopyright Stephane Guisard).}
\label{fig:mapa_campos}
  \end{center}
\end{figure}

The sample analyzed here consists of spectra for $\sim200$ stars per field that were obtained in five lines of sight toward the bulge during the first nine months of observations of the Gaia-ESO survey. In total, $1220$ stars were observed  with the ESO/VLT/FLAMES facility in the  MEDUSA mode of the GIRAFFE multi-object spectrograph, using the HR21 setting. Spectral coverage spans from $8484$  to $9001$ $\AA$ with a resolving power of $\textmd{R}\sim16200$. A total exposure time of $2700$ sec was employed, obtaining spectra with a typical signal-to-noise ratio of 90. Individual signal-to-noise ratio values and other general information are presented in Table \ref{tab:caract_campos}. In this table and in the following, fields are referred to by their names, which are assembled from their Galactic longitude (l) and latitude (b), with the p/m letter coding the $+/-$ sign: for example, the field in Baade's Window at $\rm l=+1$ and $\rm b=-4$~deg is named p1m4. A finding chart of the observed fields is provided in Fig. \ref{fig:mapa_campos}. 

The GES photometric selection was made individually for each field using J and K$_s$ magnitudes available from the VISTA Variables in The Via Lactea (VVV) project \citep{Minniti_vvv}.
A color cut was kept fixed at $(\textmd{J-K}_s)_0>0.38$. It was chosen  blue enough to not exclude the RC metal-poor giants located toward the intermediate region between the RC and the disk main-sequence (MS) plume in the CMD. The cut was reddened in each field according to the reddening estimated by averaging in a box of $20x20$ arcmin. To do this, we used the BEAM calculator\footnote{\url{http://mill.astro.puc.cl/BEAM/calculator.php}}, which is based on the reddening maps of \citet{oscarMapas_extincion}. This adjusted cut includes a field-to-field variable amount of dwarf main-sequence stars, because the reddening of the disk main-sequence plume is different from that of the bulge by a different amount in each field. The fields with more MS contamination provide the opportunity of using them as a science verification of the stellar parameter determination procedure.

 \begin{figure*}
  \begin{center}
\includegraphics[width=18cm]{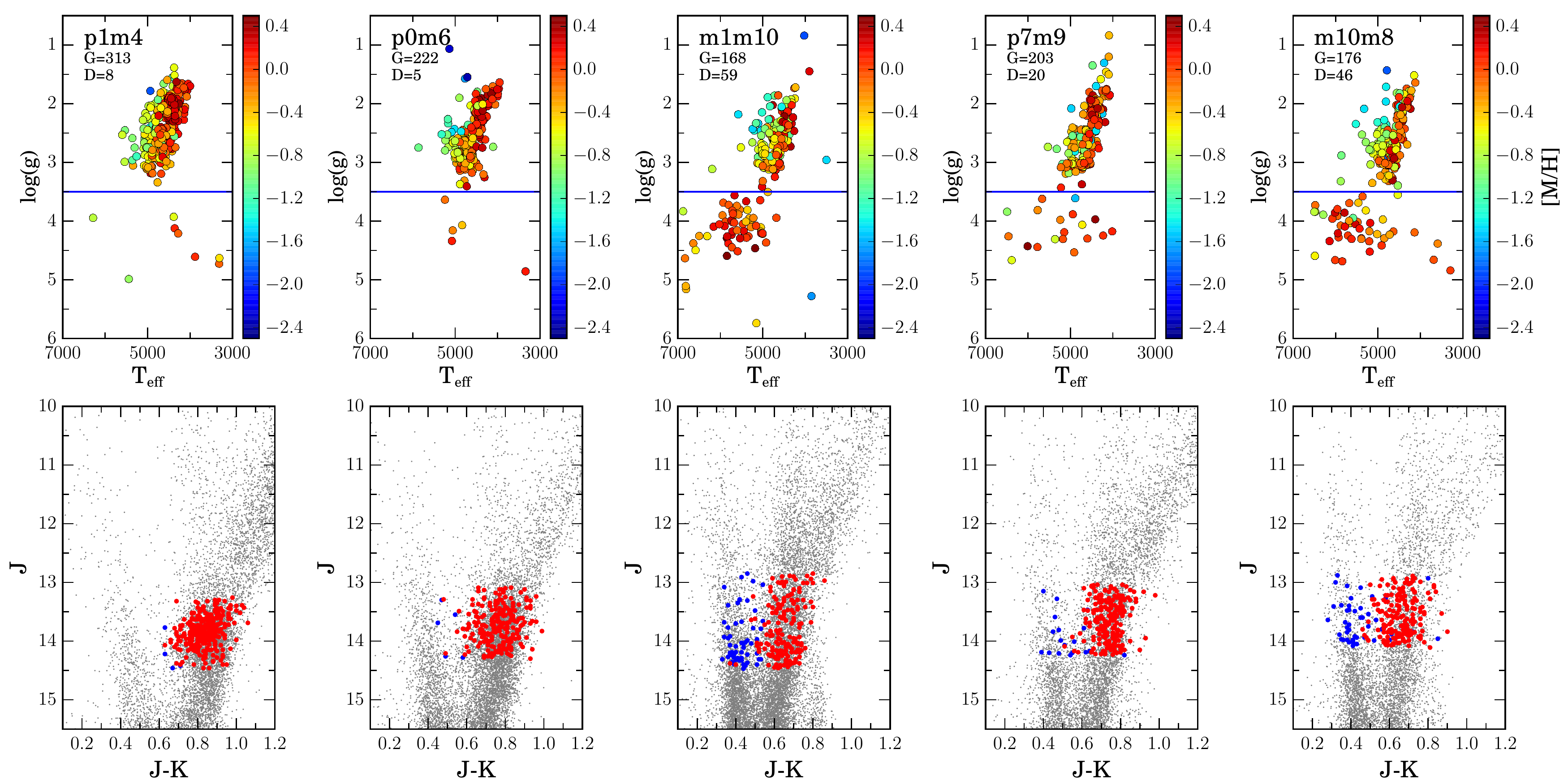}
\caption{\textit{Upper panels}: HR diagrams for the five bulge fields constructed from the stellar parameters obtained as described in the text, and color-coded according to the metallicity \met. Our adopted limit for giants (\logg$~\leq~$3.5 dex) is marked by a horizontal blue solid line in each panel. \textit{Lower panels}: J vs. (J-K$_s$) color-magnitude diagrams constructed from VVV VISTA photometry, completed with 2MASS for stars brighter than $\textmd{K}_s=12$ mag, where VVV magnitudes start to saturate. The different selected samples are apparent from the distribution of observations (colored points) in the bulge RC and MS regions in each field, color-coded according to the stellar gravity as dwarfs (blue) and giants (red).}
\label{fig:campos_ges_bulbo}
  \end{center}
\end{figure*}

The magnitude selection was generically \mbox{$(14.1-1.2)<\textmd{J}_0<14.1$}. An interval of 1.2 mag was considered to be large  enough to account for the distance spread of the bar and the changing of the mean magnitude of the RC with longitude because of the bar position angle. In practice, it was obvious from the CMD in some cases (e.g., where the RC is double)  that the faint RC would not fit in this cut. In these cases, the cut was modified to include up to 30\% of the targets in another 0.3 mag below the nominal cut.
The lower panels of Fig. \ref{fig:campos_ges_bulbo} show the combined VVV+2MASS J vs. (J-K$_s$) CMDs for the fields, with the observed samples overplotted in red and blue (giants and dwarfs). The different sampling of giants and MS stars among the fields is evident.

For calibration purposes, a subsample of $\sim110$ RC stars analyzed in \citet{hill2011} was taken from the ESO archive and processed with the complete GES bulge sample. These spectra were observed in HR21, while the original ones of \citet{hill2011} were observed in HR13 and HR14. We added these stars with the HR21-based parameters to our p1m4 sample, making it the largest one with $\sim320$ stars.

Radial velocities were determined for each star through a dedicated pipeline by cross-correlation against real and synthetic templates, as described in Koposov et al. (in prep.). Typical internal errors are on the order of 0.2 \kms.  The recommended parameters \teff, \logg, and global metallicity \met for bulge data in the internal Data Release 1 (hereafter iDR1) are those determined with the MATISSE algorithm \citep{Matisse} by the GES work package in charge of the GIRAFFE spectrum analysis for FGK-type stars. No other methods were applied to these data for iDR1, therefore the MATISSE parameters were adopted for our analysis\footnote{In the forthcoming data releases, recommended parameters are expected to be determined from homogenization of independent determinations obtained by applying several pipelines to the same dataset.}.
The processing with MATISSE uses a comparison set of synthetic spectra that are described in \citet{grid_ambre}.

To correct the derived atmospheric parameters from possible bias and set the zero point of our metallicity scale, we used a set of benchmark stars selected by the GES consortium \citep{estrellas_benchmark}. We constructed relations between the nominal atmospheric values for the benchmark stars versus those estimated in the same way as for the bulge sample stars. The obtained relations enabled us to move our original parameters onto the benchmarks scale (Recio-Blanco et al. in prep.).
The corresponding nominal errors adopted for these corrected quantities are $\Delta \textmd{T}_{\textmd{eff}}=100$ K, $\Delta \textmd{log(g)}=0.2$ dex, and $\Delta \textmd{[M/H]}=0.2$ dex (Recio-Blanco et al. in prep.).

According to these initial photometric selection criteria for the target stars, the observed samples are expected to consist mainly of: i) bulge RC and first-ascent giant stars, ii) a small contribution from foreground inner-disk RC stars, and iii), a field-to-field variable amount of contaminants from the dwarf main-sequence. The dwarf sample is clearly separated in the \teff vs. \logg plane after the atmospheric parameters are determined for the full sample (see Fig. \ref{fig:campos_ges_bulbo}). The last column of Table \ref{tab:caract_campos} shows the remaining percentages of giant stars in each field after removing the dwarf foreground stars by adopting a threshold value of \logg$\leq3.5$ dex. The upper panels of Fig. \ref{fig:campos_ges_bulbo} show the Hertzsprung-Russell diagram of each field. The metallicity-based color-code enables verifying the consistency of the parameters, with an apparent metallicity gradient with T$_{\rm eff}$, expected from isochrones for red giants. The separation between dwarfs and giants is also clear in each field, consistent with our threshold value of \logg$=3.5$, as shown by the blue horizontal lines. It is also consistent with the amount of stars located in the disk MS region (bluer than the RC location) of the  corresponding CMDs in the lower panels. These dwarf stars have, on average, a metallicity close to solar, and a lower radial velocity dispersion than giants in each field. These properties characterize them as disk stars distributed along the line of sight.


\section{Reddening determinations}
\label{sec:determinacion_reddenings}
By comparing our purely spectroscopic parameters with photometric parameters, we can estimate the extinction in each star's line of sight, which in turn enables us to perform a consistency check for the atmospheric parameters presented in the previous section.
After the spectroscopic cleaning of the sample from dwarf stars, the final data set consists mainly of bulge RC stars plus a contribution from internal disk RC and a small fraction of halo giant stars. In this way, most of the stars are confined in the bulge region. Their reddening values are then, in essence, determined by the extinction through the disk, where the observed dust that obscures the bulge region is located. For this reason, we can expect, except for differential reddening, similar extinction values for stars belonging to the same field.
In the following, we describe how we estimated reddening values from the spectroscopic parameters.

We can estimate the reddening toward a star by comparing the color from photometry (affected by reddening) with the spectroscopic (reddening-free) temperature transformed into a color by adopting an empirical color calibration. The ancillary VISTA VVV photometry provides the apparent J-K$_s$ color for each star. To compute the extinction, we need to determine the true J-K$_s$ color inferred from the atmospheric parameters of the star. We made use of the empirical calibration of \citet{calib_bonifacio}, which was constructed using the infrared flux method on the basis of 2MASS photometry. First of all, we used the following transformation between 2MASS colors and J-K$_s$ color in VISTA system, from the VISTA VVV data release 1.1\footnote{\url{http://www.eso.org/sci/observing/phase3/data_releases/vvv_dr1.html}}
\begin{eqnarray}
J_{VVV}&=&J_{2MASS}-0.077(J_{2MASS}-H_{2MASS}),\\
H_{VVV}&=&H_{2MASS}+0.032(J_{2MASS}-H_{2MASS}),\\
K_{VVV}&=&K_{2MASS}+0.010(J_{2MASS}-K_{2MASS}).
\end{eqnarray}
to construct a J-K$_s$ color in the VISTA system. In this relation, the necessary spectroscopically based colors in the 2MASS system are estimated by inverting  equation (10) of \citet{calib_bonifacio} and keeping only the physically meaningful solutions. By using the true color estimated in this way, we obtain the reddening from E(J-K)=$(J-K)_{\textmd{photo}}-(J-K)_{\textmd{inferred}}$.

While applying the empiric calibration, we kept only the solutions inside their range of applicability to avoid values in extrapolation. In addition, to obtain reliable extinction values for most of the stars in our sample, we selected stars with $0.9\leq$\logg$\leq3.5$ dex, \met$\leq0.2$ dex, and \teff$\geq3500$ K. In this way, we avoided the coldest giants for which the atmospheric parameter determination might be more uncertain. We attributed the average values of their parent fields as individual extinction for the excluded stars (232 of $1~082$ giants).

We estimated errors for the determined extinctions as follows: we adopted $\Delta$T$_{\textmd{eff}}=100$ K and $\Delta$[M/H]$=0.2$ dex as nominal errors in the spectroscopic parameters (see Sec.~\ref{sec:datos}). The photometric error of VVV data at the position of the RC in a crowded bulge field is $\lesssim0.01$ mag in the J band and $\sim0.025$ mag in the K$_s$ band \citep{vvv_dr1}. Therefore, we did not consider the errors in J for the computations and just adopted a generic value of 0.025 in K$_s$. For each star, we performed $1000$ Monte Carlo realizations of \teff, \met, and K$_s$. The values were generated randomly from Gaussian distributions centered on the respective  values, with a dispersion according to the nominal errors in each quantity. For these random sets of parameters, we calculated the associated extinction, and the final error estimate for the star was taken as their standard deviation. We obtained typical errors of $\sim0.03$ mag on E(J-K).

\begin{table} 
\centering
\caption{Reddening values for the five program fields. We compare our spectroscopic E(J-K) determinations (spec) with those provided by \cite{oscarMapas_extincion} (G11) and \cite{schlegelMapas} (S98). Dispersion values for our estimates come from the valid values, after an MAD-based clipping (red points in Fig. \ref{fig:distro_extincion_campos}). Dispersion values $\sigma_{phot}$, associated with  the determinations of G11, come from the procedure explained in the main text.}
\begin{tabular}{l|rr|rrr}
\hline
\hline
Field name & spec & $\sigma_{spec}$ & G11 &$\sigma_{phot}$ & S98   \\\hline
p1m4  & 0.21 & 0.04 & 0.25 & 0.025 & 0.34 \\ 
p0m6  & 0.14 & 0.05 & 0.15 & 0.009 &0.25 \\
m1m10 & 0.06 & 0.03 & 0.03 & 0.009 &0.07 \\
p7m9  & 0.11 & 0.05 & 0.10 & 0.021 &0.18 \\
m10m8 & 0.06 & 0.04 & 0.02 & 0.008 &0.11 \\ \hline
\end{tabular}
\label{tab:extinction_campos}
\end{table}

A cut based on the median absolute deviation (MAD) was applied to remove extreme values from the set of extinctions computed in each field. From these cleaned sets, we estimated the average and dispersion values reported in Table \ref{tab:extinction_campos}. The dispersion ($\sigma_{spec}$) is small in each field and traces a combination of the propagated error in the individual stellar parameters and the differential extinction within each field, and may also be affected by inner-disk star contaminants (with slightly lower extinction values). To estimate the impact of differential extinction, we computed extinction values in a spatial grid in steps of 5 arcmin, centered on each of our fields and spanning an area equivalent to a GIRAFFE field of view (25x25 arcmin$^2$). We used the extinction maps of \citet{oscarMapas_extincion} from the BEAM calculator. The corresponding  dispersion for each field is quoted as $\sigma_{phot}$ in Table \ref{tab:extinction_campos}. These estimates provide an upper limit for the differential reddening in each field and are systematically lower than the dispersions ($\sigma_{spec}$) among our spectroscopic determinations, which shows that the error due to the stellar parameter error propagation ($\sigma_{params}$) dominates the total dispersion $\sigma_{spec}$. 
Assuming that the internal errors in G11 are small and that $\sigma_{phot}$ is a fair indication of the true differential extinction, we can set an upper limit for the error on E(J-K) due to the stellar parameter error propagation $\sigma_{params}$ from 0.029 to 0.049 mag in our fields, in close agreement with the mean error estimate provided in the previous paragraph.

In addition, Table \ref{tab:extinction_campos} compares the average reddening obtained in each field (spec) with that provided by \cite{oscarMapas_extincion} (G11) and by \cite{schlegelMapas} (S98). The G11 map was computed using the RC location in VVV photometry as a proxy for reddening, while the S98 values are in a region with known problems due to high extinction and small-scale variations. Additionally, because S98 provided integrated extinctions, it is expected that they  provide upper limits to the extinction of individual stars along the line of sight. Our values are compatible with those of G11 and are always lower than those of S98, which are well known to be overestimated in low-latitude fields. From the small size of the derived internal dispersion and the successful comparison with G11, we conclude that our stellar parameters are robust and consistent within the errors.

\begin{figure}
  \begin{center}
  \includegraphics[width=9cm]{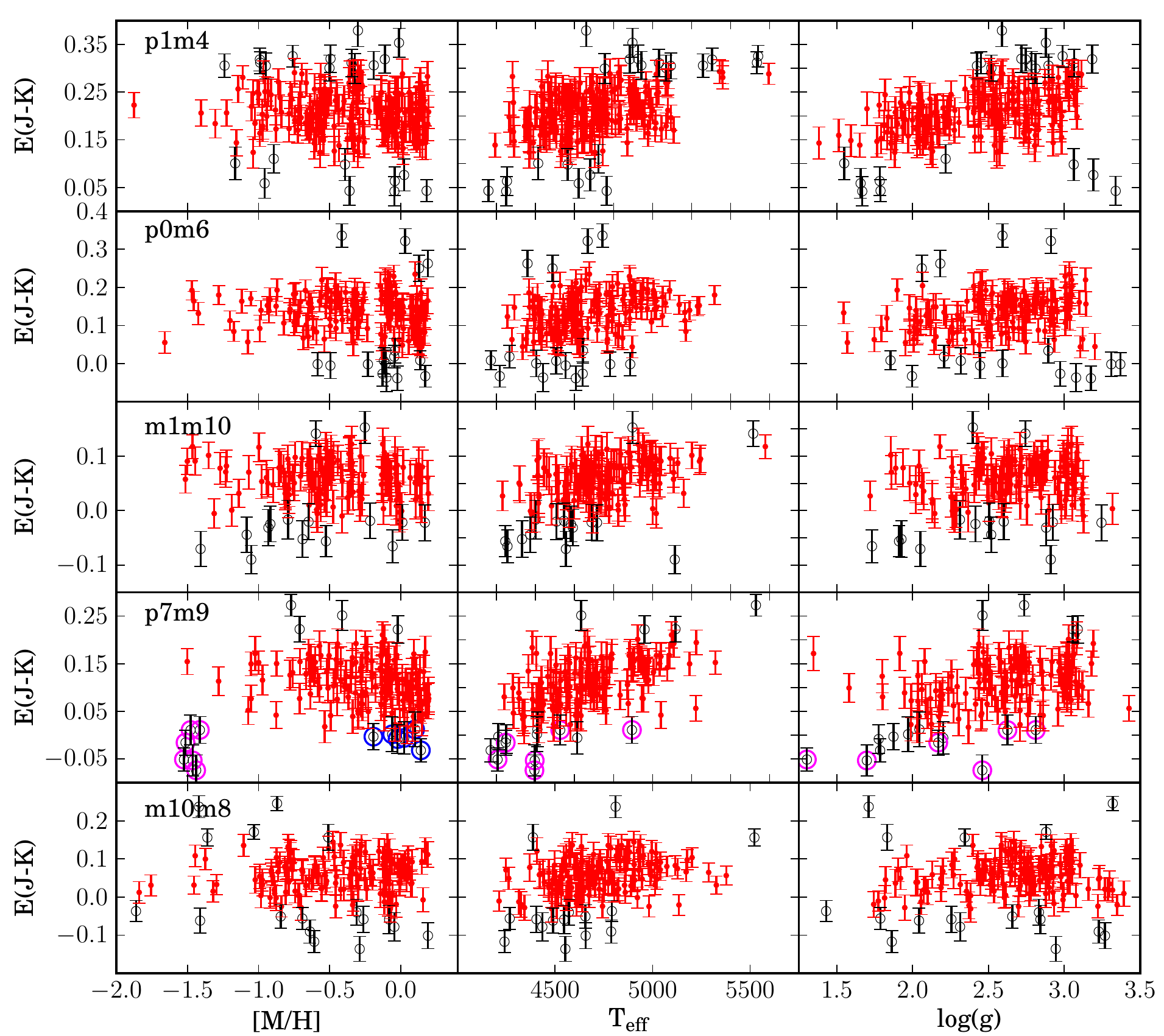}
\caption{Distribution of estimated reddening values for each field (rows) with respect to the atmospheric parameters (columns). Red points and black circles represent values selected and clipped using median absolute deviation statistics. Error bars are computed using Monte Carlo simulations, as described in the main text. In the p7m9 field, blue points mark the high-metallicity bulk of clipped points, while magenta highlights stars clumped out of the general trend of this field.}
\label{fig:distro_extincion_campos}
  \end{center}
\end{figure}

In Fig. \ref{fig:distro_extincion_campos}, we display reddening values versus atmospheric parameters. In general, there are no strong trends of reddening with metallicity. There are some weak trends with \teff and \logg that can be explained by taking into account the known correlation between the errors on those parameters. There is indeed an incompletely broken degeneracy when using the infrared Calcium triplet region (HR21 setup) to estimate stellar parameters \citep{kordopatis_matisse}. Even with these trends, we can verify the general consistency of the determinations, with smooth distributions that show a small dispersion, as quoted in Table \ref{tab:extinction_campos}.

On the other hand, we examined the p7m9 field with particular care because, as shown in the next section, it presents a strong metal-rich peak in its metallicity distribution function. Some of the points clipped in the [M/H] vs. E(J-K) plane of p7m9 (blue circles in Fig. \ref{fig:distro_extincion_campos}) correspond to an overdensity with high-metallicity values and low extinctions. In general, the bulk of points located at this position in the diagram can be interpreted as stars located closer to the Sun. This could be consistent with stars that are located in the inner disk, with a metallicity around solar and lower extinctions due to the shorter line of sight.
In the same field, there is a clump of points with low extinction at \met$\sim-1.5$ dex. These stars do not follow the general trend of the field. However, after individual examination, we verified that they do not present particular problems with the parameter estimations.

In summary, using the stellar parameters of our giant samples, we obtained extinction values consistent with those determined from independent methods. From our error analysis, we identified that the main source of uncertainty in our spectroscopic method comes from the propagation of stellar parameter errors, while the contribution from differential extinction is modest. 


\section{Final selected sample}
\label{sec:seleccion_final}
As explained in Sect. \ref{sec:datos}, the original GES photometric criteria select stars with a nominal criterion, adapted in magnitude according to the observed morphology of the luminosity distribution function. The idea is to sample stars in the complete RC region, considering the cases where the split RC overdensities are more separated. Moreover, as the reddening is different for bulge and foreground disk stars, the final samples are mainly located in the bulge RC region, with some contribution from the disk RGB plus a variable amount of disk dwarfs.
The estimated surface gravity enabled us to reject the dwarf population. On the other hand, to work on the basis of a homogeneous sampling, we selected for the subsequent analysis just the stars in the RC magnitude region in each field. 

\begin{figure}
  \begin{center}
  \includegraphics[width=8cm]{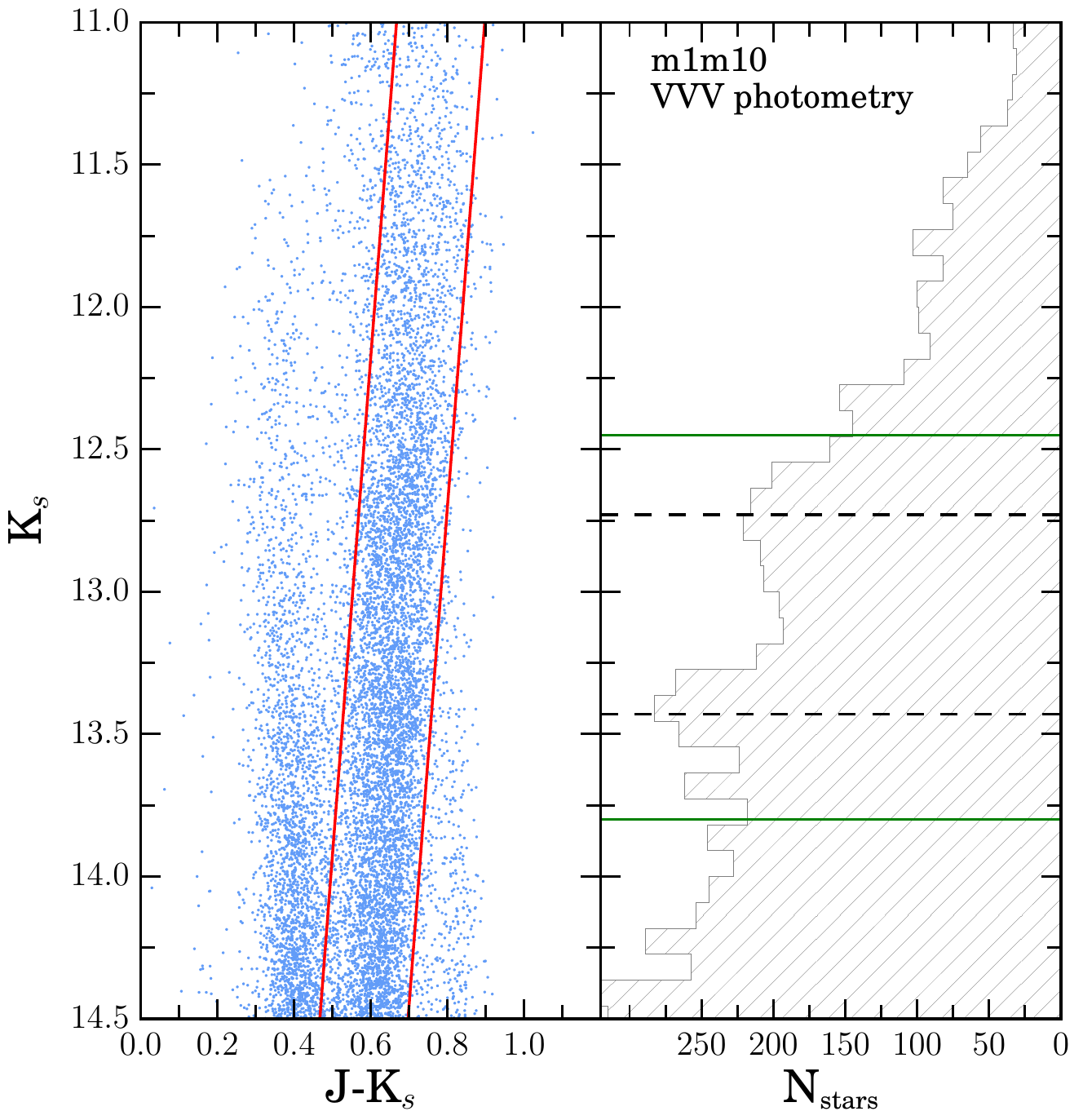}
\caption{CMD and luminosity function for the m1m10 field. This figure illustrates the procedure used to obtain magnitude limits for the RC members. The strip of stars between the red lines, following the giant sequence in the CMD, is used to construct the histogram in the right panel. The green horizontal lines mark the adopted limits for the RC, while the dotted lines mark the position of the peaks detected in the distribution. We use the estimated limits to select bona fide RC stars in the respective spectroscopic samples, and also to explore the double RC morphology in Sect. \ref{sec:rc_split}. Note that unlike the CMDs presented in Fig. \ref{fig:campos_ges_bulbo}, this contains only VVV photometry that saturates at K$_s$=12 mag. The RC region is fainter than this limit.}
\label{fig:ejemplo_rc_cut}
  \end{center}
\end{figure}

To do this, we constructed a separate luminosity function of the giant branch for every field, using the available VVV photometry without any correction for reddening. In Fig. \ref{fig:ejemplo_rc_cut}, we depict the procedure for the field m1m10. The strip of stars between the red lines in the left panel mainly corresponds to RC stars, with those at the bulge distance clustered in the region around $K_s=13$ mag. With this selection we constructed a luminosity function similar to that shown in the right panel of Fig \ref{fig:ejemplo_rc_cut}. The magnitude limits below and above the bulge RC were determined from this distribution \footnote{The limits are generically 12.45 and 13.75 mag in K$_s$, with field-to-field variations of the order of 0.05 mag.}. Two horizontal solid green lines display the limits for m1m10. Using those individually determined limits, the samples are filtered to retain only the bona fide RC bulge stars. The purpose of selecting stars using limits determined from each specific field luminosity function is to adjust the selection with respect to the maximum of the star density in the line of sight. Additionally, this selection ensures an homogeneously behaved contamination from field to field.
Some disk giant branch stars could remain in the selection because the combination of distance and absolute magnitude allow them to fall between the magnitude boundaries. We stress here that this selection is only based on a magnitude cut. A color cut was implicitly made when the dwarf stars were separated from the sample based on the \logg values. The final subsamples comprise 304 stars for p1m4, 205 for p0m6, 140  for m1m10, 203 for p7m9, and 154 for m10m8, for a total of $1006$ stars.\\

To analyze the metallicity distribution function of Sect. \ref{sec:distros_met}, we verified whether the photometric cuts   introduce a bias while removing some stars from each field. We estimated the percentage of eliminated stars in  metallicity bins spanning the complete range of values in our samples. We verified that the profiles are fairly flat with a deleted rate of about  9$\%$.\\

\begin{figure*}
 \begin{center}
  \includegraphics[width=18.2cm]{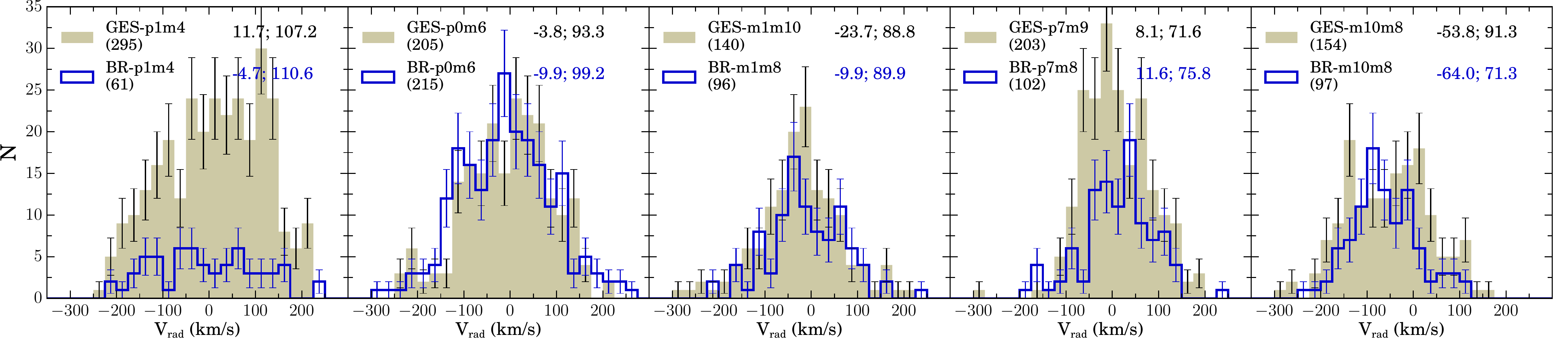}
\caption{Velocity distributions (heliocentric, in \kms) for the five program fields shown as filled histograms. We consider here only the stars kept in our final selection. For the p1m4 field, we exclude the stars identified as possible globular cluster members (see main text). For comparison, we overplot the velocity distributions of BRAVA data (blue solid line) of the corresponding or nearest available field as indicated in each box. The number of stars (in brackets) and the mean and $\sigma$ values for each group of data are also quoted.}
\label{fig:vrad_ges_bulbo}
  \end{center}
\end{figure*}

In the same context, a bias in the MDF can be introduced during observations independently of the adopted selection function. In GIRAFFE, the fiber allocation process tries to maximize the number of targets while avoiding forbidden fiber crossings. The observed stars are in this way taken randomly from a predefined list of candidates. This can introduce inhomogeneities in the color sampling. To obtain an unbiased picture of the MDF, we require a sampling homogeneity with respect to the true metallicity distribution. We determined the impact of this bias for our sample in the following way: Using the complete VVV photometric catalog, from which the observed samples were selected, we computed the fraction between observed and available stars in several color bins. In an unbiased sample, the relative amount of stars should be constant throughout the whole color interval. We checked that the computed profiles are flat within errors, with an average value of 9.6 $\%$ and a dispersion of 5.6$\%$. We judge that the small variations inside these limits are too small to introduce a significant bias in the computed metallicity distribution function of each field, even more so because the color-metallicity correlation is weak.

In conclusion, the final samples in each field are composed of stars that span the complete color range, inside the magnitude range of the respective RC overdensity. We verified their homogeneity in color sampling and the existence of potential bias introduced by adopting the magnitude cuts. They are suitable for the subsequent analysis.


\section{Radial velocity characterization}
\label{sec: caracterizacion_vel_rad}

A first check for the kinematic properties of our fields and variations over the area sampled by them can be made by analyzing their radial velocity distributions. The good quality of our data allow us to obtain radial velocities with small internal errors from the dedicated pipeline, as presented in Sect. \ref{sec:datos}.

In Fig. \ref{fig:vrad_ges_bulbo}, we present the radial velocity distributions of the selected RC stars in our fields compared with those obtained by the BRAVA (M giants) project \citep{kunder_brava}. In each case, we selected for comparison the nearest BRAVA field from the available data set (cf. Fig. \ref{fig:mapa_campos}). In general, our distributions resemble those of BRAVA and are moreover based on larger samples.  
For each field, the global distribution shape, the mean, and the dispersion are consistent, even for the fields where the BRAVA comparison field is not exactly at the same location ($\Delta b=2^\circ$ for m1m10, which is the largest difference with respect to a BRAVA comparison field). Two of the fields present some peculiarities, however: m10m8 and p1m4. 

The p1m4 velocity distribution presents two peaks, but one of them disappears after filtering for possible contaminants from the globular cluster NGC 6522 ($V_{\textmd{rad}}=-21.1\pm3.4\textmd{ km }\textmd{s}^{-1}$, \feh=-1.34 dex, R$_\textmd{H}=1$ arcmin, \citeauthor{harris} \citeyear{harris}), which is located close to the center of this field. In this way, we removed nine stars from the p1m4 final sample (already removed from the respective panel in Fig. \ref{fig:vrad_ges_bulbo}). We suspect that the second feature at $\sim120$ \kms might only be an artifact from our small-number statistics. The stars belonging to this peak do not seem to share common properties suggest that they form a substructure.

We performed a two sample Kolmogorov-Smirnov (K-S) test to compare our velocity distributions with those from BRAVA. The GES and BRAVA distributions in p1m4, p0m6, p7m9, and m1m10 are compatible, with a confidence (\textit{p-value}) of $25\%$, $36\%$,  $36\%$, and $48\%$, respectively.

On the other hand, the m10m8 distribution presents a secondary narrow peak at $\sim-140\textmd{ km }\textmd{s}^{-1}$. The stars in this peak apparently only share the radial velocity, therefore we do not rule out that it might be an artifact from small-number statistics. The K-S test value of $7\%$ shows that the differences observed are still compatible with the two samples coming from the same parent distribution. 

We point out here that in principle, we do not expect to obtain an absolute agreement between the GES an BRAVA radial velocity distributions because of the different probes and observational strategies adopted in the two projects. The Gaia-ESO survey samples the Galactic bulge using RC stars. As standard candles, RC stars cluster in color magnitude diagrams when the sampled population is also spatially clustered. For a standard candle, a spread in distance of $\pm3$ kpc around the bulge center at 8kpc implies a difference of $\sim1.7$ magnitudes, which is large compared with the typical magnitude size of the RC in our fields.

In particular, this amount is larger than the typical separation between the splitted RCs present in some of our fields (cf. Sect. \ref{sec:rc_split}), which corresponds to 0.6 magnitudes. In fact, when the photometric selection cuts stars around the RC magnitude center, we are implicitly cutting in distance. Therefore, the RC stars are mainly probing a region in each field where the RC stars are spatially clumped, which is the central part of the bulge. This is not necessarily the case for the BRAVA sample of M giants. Because these stars are not standard candles, it is not possible to determine their spatial distribution. The limits in the BRAVA magnitude selection in a 1.1 magnitude range \citep{Howard_brava_selection} ensure a relatively contamination-free sample of bulge stars without any bias to oversample a specific region. They are probably homogeneously distributed across the bulge region and only follow the general density properties, which means that they trace every structure there, bar, internal disk, and/or some classical spheroid component.


\section{Metallicity distribution functions}
\label{sec:distros_met}
The analysis of the  metallicity distribution functions and their ($l$,$b$) dependence is a powerful tool for unveiling the complex bulge structure and possible stellar population mix.
As explained in Sect. \ref{sec:seleccion_final}, we selected a sample of stars centered on the red clump, including an unavoidable amount of RGB disk contaminants.

\begin{figure*}
  \begin{center}
\includegraphics[width=17.4cm]{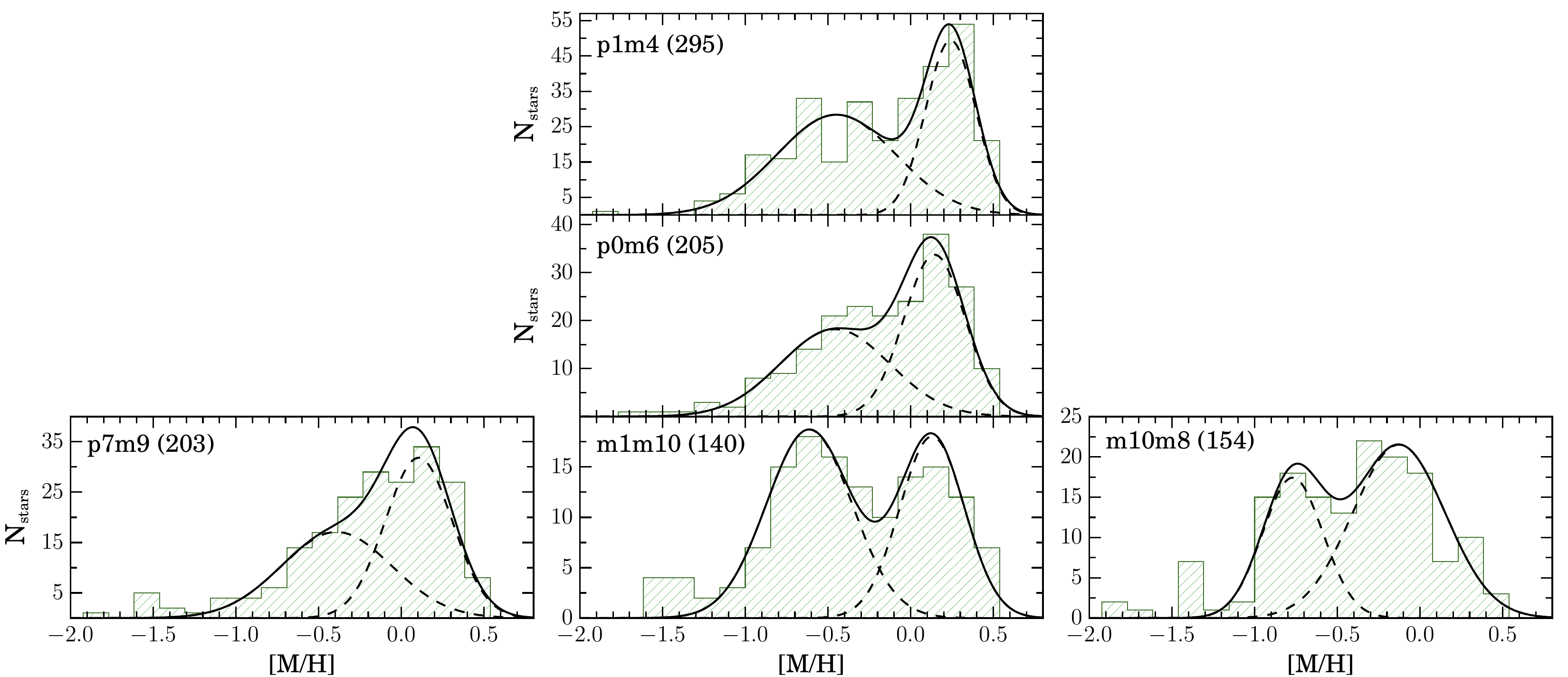}
\caption{Metallicity distribution functions for the five bulge fields, presented as dashed color histograms. Only stars selected as RC bulge members are considered. Black dashed and solid lines show the components and the global fitted model. The three minor-axis fields p1m4, p0m6, and m1m10 are shown in the central panels, and the lateral fields p7m9 and m10m8 at the left and right.}
\label{fig:idf_ges_bulbo_hist}
  \end{center}
\end{figure*}

\begin{figure*}
  \begin{center}
\includegraphics[width=16.8cm]{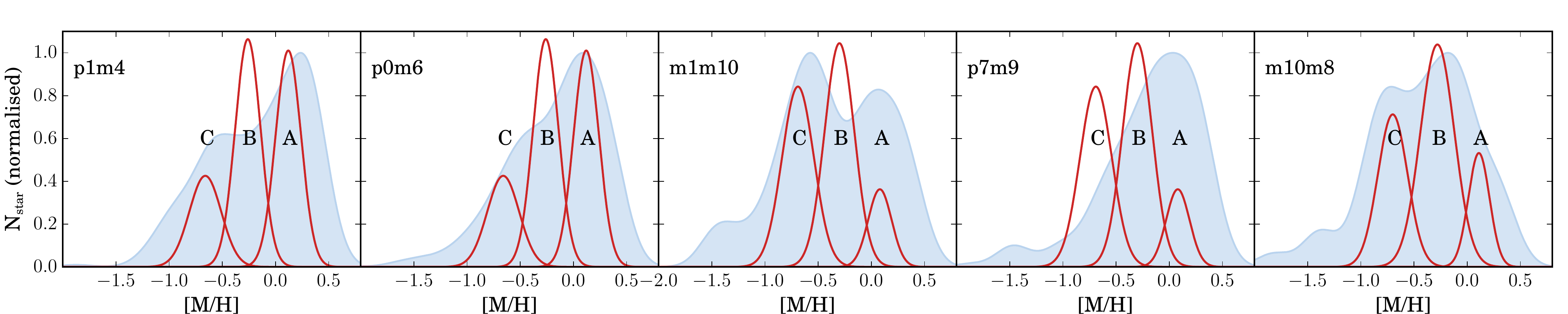}
\caption{Metallicity distribution function for each field presented as generalized histograms (filled profiles) using a smooth parameter of 0.15 dex. We use generalized histograms to show the data structure and avoid the effect of arbitrary bin limits in conventional histograms. Overplotted (red solid lines) are the populations reported by  \citet{argos_3} for the nearest corresponding field.}
\label{fig:idf_ges_bulbo_hist_gen}
  \end{center}
\end{figure*}

Figures \ref{fig:idf_ges_bulbo_hist} and \ref{fig:idf_ges_bulbo_hist_gen} show the MDF for the observed fields.
The MDFs show some substructure, which suggests a complex nature. We decomposed them using a number of Gaussian profiles. To do this, we adopted a Gaussian-mixture models (GMM) formalism from the python package \textit{sklearn}. We tried fits with 1 up to 4 components, using the expectation-maximization algorithm to maximize the likelihood function of every Gaussian-mixture combination. To decide which was the best model fit with the number of components supported by the data and their relative weights, we used the Akaike information criterion (AIC). This provides a way to perform a model selection, which is a measure of the relative  quality of a statistical model proposed to fit a given data set. As a relative quantity, the AIC does not provide information about the quality of the best model in an absolute sense. Instead, it tries to find the best fit using a penalty proportional to the number of estimated parameters, to avoid overfitted solutions.\\
Figure \ref{fig:idf_ges_bulbo_hist} displays the components identified in each field as dashed Gaussian functions, with the sum as a solid line. We performed the fit using stars with \met$>-1.3$ dex to avoid the undersampled metal-poor tails (see Sect. \ref{subsec:mdf_metal_poor_stars} for the analysis of the metal-poor end stars).\\ 


In summary, we can identify a mixture between two populations in each field, a narrow metal-rich population, hereafter called (i) and a metal-poor one (ii), which is in general broader than the former. The exact position of the two peaks varies from field to field, but the first peak is always at supersolar metallicity and the second appears at subsolar metallicity, except for m10m8, the most metal-poor MDF among our fields.
Table \ref{tab:mezcla_gausianas} summarizes the characteristics of the fitted models and lists the mean metallicity, dispersion, and relative weights of each component.\\
The relatively limited statistics of our individual field samples has to be kept in mind, however.  The number of stars currently observed in each field is not large enough to explore a more detailed decomposition with a higher number of components, as in the analysis of \citet{argos_3} (cf. Sect. \ref{subsec:mets_comparacion_literatura}). However, our analysis has the advantage of comparing specific individual fields, while \citet{argos_3} merged stars in very wide strips in $l$, which may have blurred some intrinsic gradient in their components.

\begin{table} 
\centering
\caption{Characteristics of the different MDF components in each field. We quote in brackets the average radial projected distance in degrees from $(l,b)=(0,0)$ for each field.}
\begin{tabular}{cccc}
\hline
\hline
Name (distance deg)   & Mean \met & $\sigma$ & weight \\\hline
p1m4 (4.12) &  $-0.16\pm0.03$ &  -- & --    \\
(i)   &  0.24  & 0.15 & 0.42 \\
(ii)  & -0.45  & 0.36 & 0.58 \\  \hline
p0m6 (6.0) & $-0.17\pm0.03$  &  --  & --  \\
(i)   &  0.14  & 0.19 & 0.51 \\
(ii)  & -0.46  & 0.33 & 0.49 \\  \hline
m1m10 (10.05) & $-0.37\pm0.04$ & --   & --  \\
(i)   &  0.13  & 0.20 & 0.42 \\
(ii)  & -0.61  & 0.26 & 0.58 \\  \hline
p7m9 (11.35) & $-0.18\pm0.03$ & --   & --  \\
(i)   &  0.11  & 0.20 & 0.53 \\
(ii)  & -0.39  & 0.33 & 0.47 \\   \hline
m10m8 (12.73) & $-0.41\pm0.04$ & --  & --  \\
(i)   & -0.13  & 0.28 & 0.65 \\
(ii)  & -0.77  & 0.18 & 0.35 \\  \hline 
\end{tabular}
\label{tab:mezcla_gausianas}
\end{table}

\subsection{Vertical gradient and positional variations of the MDFs}
\label{subsec:gradiente_vertical}
By comparing fields on the minor axis, \citet{zoccali2008} found a metallicity gradient of $\sim-$0.075 dex/deg, while \citet{gonzalez_metallicity_map} measured a gradient twice smaller of $\sim-$0.04 dex/deg from their photometric metallicity map. From the p1m4, p0m6, and m1m10 fields, we explored the presence and properties of such average metallicity variations on the bulge minor axis.\\
The average metallicity values for p1m4 and p0m6 are the same ($-0.16\pm0.03$ and $-0.17\pm0.03$ dex, respectively), both are higher than the value of $-0.37\pm0.04$ dex for m1m10. This is compatible with the metallicity map of \citet{gonzalez_metallicity_map}, where a strong change in the gradient is seen at $b\sim-6^{\circ}$, while the internal region appears to be more uniform. In our data, the MDF average value decreases from p0m6 to m1m10, which establishes a gradient of -0.05 dex/deg.\\
This change in mean metallicity is caused by the variation in the relative weight of components (i) and (ii), with a larger proportion of metal-rich stars confined to the central regions of the bulge.\\
The main effect of moving outward on the bulge minor axis is the decrease of the contribution of component (i) to the global MDF. Our work is based on a RC sample, while the sample of \citet{zoccali2008} is composed of K giants that are located $\sim1$ mag above the position of the RC.\\
Within errors, there are no significant differences between the general shapes of the MDFs of p1m4 and p0m6. Both distributions have a pronounced metal-rich peak (component (i) containing the majority of the stars), and a less significant but wider component (ii). The relative proportion of metal-poor stars increases noticeably in the m1m10 field. This field presents well-defined and more detached components, with a prominent component (ii) and a smaller but still significant component (i).\\
Finally, the mean position of component (i) presents small variations along the minor axis (Table \ref{tab:mezcla_gausianas}), but still well inside the generic parameter error of 0.2 dex (Sect. \ref{sec:datos}). On the other hand, the mean position of component (ii) is similar for the two internal fields p1m4 and p0m6, but different from that of the m1m10 field. The difference in the mean metallicity of component (ii) between the two inner and the outer field is $\sim 0.2$dex, this time similar to or larger than the uncertainties. This may be a hint that component (ii) hosts a real metallicity gradient with |b|.

\subsection{Off-axis fields}
\label{subsec:off_axis_fields}
Two of the five observed fields are located away from the minor axis, at similar latitudes of $\rm b\sim -9 \pm 1$ deg, and opposite longitudes.

Interestingly, the p7m9 field lies in a highly metal-rich region in the map of \citet{gonzalez_metallicity_map}, unlike the symmetric fields at negative longitudes, which at these latitudes far from the Galactic midplane are rather metal-poor. Furthermore, the density maps of \citet{wegg_mapas} show that the p7m9 line of sight intersects the close branch of the  X-shaped bar. The general shape of the p7m9 MDF, is consistent with a line of sight intersecting a metal-rich region. In fact, the MDF has a prominent peak at \met$\sim0.10$ dex and a relatively sharp tail decreasing to lower metallicity. Our Gaussian decomposition leads to two components, with the metal-rich being narrow and stronger than the metal-poor component. The average field metallicity (-0.18 dex) is similar to the p1m4 and p0m6 fields.

On the other hand, the bulk of stars in m10m8 have low metallicity, in agreement with what is expected for a field far for the midplane and from the minor axis. Its mean metallicity of -0.41 dex, in agreement with \citet{gonzalez_metallicity_map}, confirms this general picture.
The m10m8 MDF is decomposed into two populations of similar sizes. Both components are more metal-poor than the respective component in other fields. The component (i), already peaked at subsolar metallicity, is moderately wider than the metal-rich components of the other fields (Table \ref{tab:mezcla_gausianas}). The component (ii), in contrast, appears to be slightly narrower. It peaks at -0.77 dex, the lowest value in the field samples.

The contrasting properties of p7m9 and m10m8 MDFs are consistent with the observed asymmetry in the map of \citet{gonzalez_metallicity_map} with respect to the minor axis. The asymmetry was attributed in \citet{gonzalez_metallicity_map} to a perspective effect. Since at positive (resp. negative) longitude, the line of sight intersects the near (resp. far) sides of the bar, the stars at positive longitudes are, on average, closer to the observer and hence to the plane than those at negative longitudes and same $b$. This fact, together with an intrinsic bulge vertical metallicity gradient, could explain the observed metallicity  asymmetries. On the other hand, 
we have shown in the previous section that the gradient through the minor axis can be explained from a variable mix ratio of metal-poor and metal-rich stars with $b$ and not a solid shift-type of metallicity gradient. In this context, the MDFs of the off axis fields can be interpreted in a slightly different way, where the metallicity components (i) and (ii) would correspond to separated structures with different spatial distributions, which allows reproducing the map asymmetry from their particular geometrical properties. \\
In summary, we have shown in this section that the MDF is decomposed into two metallicity groups, whose relative proportion accounts for the mean metallicity variations in the bulge region. The metal-rich component (i) is a narrow distribution with a nearly constant mean metallicity in our fields, except for m10m8, where it is at subsolar metallicity. The metal-poor component (ii) is wider and on average more metal-poor with increasing ($l$,$b$),  with the exception of p7m9, where the component is less metal-poor than expected for its location.

 \subsection{Metal-poor stars}
\label{subsec:mdf_metal_poor_stars}
\begin{figure}
  \begin{center}
\includegraphics[width=9cm]{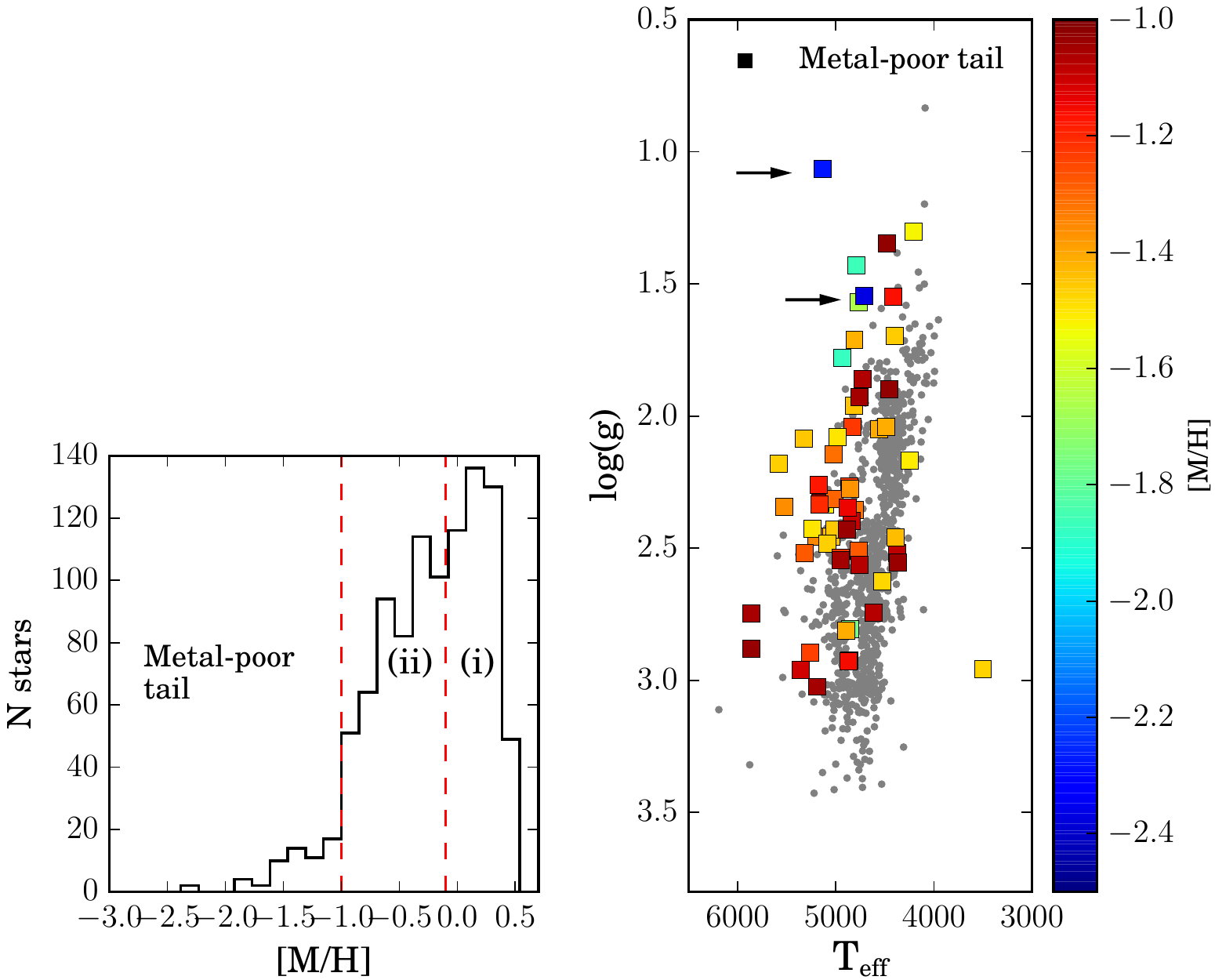}
\caption{\textit{Left panel: }Global MDF obtained by merging the complete studied sample. Two vertical dotted red lines split the metallicity range spanned by the distribution into three regions, the two metallicity components (i) and (ii), and the metal poor tail of stars with \met$\leq-1.0$ dex. \textit{Right panel: }HR diagram for the complete sample. Stars located in the metal-poor tail of the MDF (\met < -1.0 dex) are highlighted with  a color code according to their metallicity. The two most metal-poor stars of our sample are marked by arrows.}
\label{fig:mdf_metal_poor}
  \end{center}
\end{figure}

An important matter while analyzing the MDF is to explore its metal-poor end to compare it with the  predictions from formation models. To study this poorly sampled regime in our fields, we have improved statistics in the metal poor tail by merging the complete sample. This allowed us to roughly estimate the percentage of stars in the metal-poor bins. Nevertheless, we warn here that while exploring the bulge MDF using red clump stars, a possible cut in the metal-poor end could be introduced by stellar evolutionary effects. With decreasing metallicity, stars eventually undergo the helium burning phase in the horizontal branch instead of in the RC. The exact metallicity limit depends on very poorly modeled processes such as the mass-loss rate, which prevents a precise estimate from theoretical calculations. On the other hand, we already verified that the photometric selection limits imposed in Sect. \ref{sec:seleccion_final} do not bias the MDF against the more metal-poor stars because we expect to include first-ascent RGB stars in the sample as well.\\
In view of the differences between the individual field MDFs, we expect to blur the shape of the metallicity components (i) and (ii)  when we merge the complete bulge sample, but we favor the definition of the metal-poor end. The global MDF is presented in the left panel of Fig. \ref{fig:mdf_metal_poor}. The bulge MDF extends continuously until $\sim-1.9$ dex. After a gap, two stars populate a single bin at $-2.3$~dex. The right panel presents the HR diagram of the stars in the metal-poor tail overplotted with the complete sample. The color code shows the metallicity of the metal-poor stars with \met$\leq-1.0$ dex. In this diagram, the two stars in the bin at $\sim-2.3$ dex are stars with low \logg values, located at the top of the distribution. The star with the lowest surface gravity should be placed at a larger distance to be compatible with our photometric selection, and consequently is most probably a halo star. The second star has a gravity still compatible with a location in the bulge region (roughly the region inside $R_G<3.5$ kpc, \citeauthor{argos_3} \citeyear{argos_3}).

In conclusion, if we consider the metal-poor end of the continuous MDF in Fig. \ref{fig:mdf_metal_poor} (e.g, excluding the bin at$\sim-2.3$ dex), the cut occurs at -1.88 dex (two most metal-poor stars in the continuous MDF regime). Our sample shows that the incidence of stars with \met $<-1.0$ dex is 5.8\% and only  1.1\% for stars with \met$<-1.5$ dex (for comparison, the respective values from \citeauthor{argos_3} \citeyear{argos_3} are 4.4\% and 0.7\%). This low ratio stresses the importance of constructing large samples to increase the statistical significance of the metal-poor tail.

\subsection{Comparison with literature studies}
\label{subsec:mets_comparacion_literatura}

Figure \ref{fig:idf_ges_bulbo_hist_gen} shows as filled profiles the generalized histograms of the MDFs, using a smoothing kernel of 0.15 dex. The MDF components of \citet{argos_3} are also overplotted as red Gaussians. For each field, we  associated the closest of the three $b=-5^\circ$, $-7.5^\circ$ or $-10^\circ$ ARGOS stripes. We recall that in the study of \citet{argos_3} the three MDFs used for this comparison were constructed by merging several fields at the specific latitudes that covered $\pm15^\circ$ in longitude. Bearing in mind the global picture of the metallicity map of \citet{gonzalez_metallicity_map}, the mixture of fields across that longitude range implies the risk of combining fields with different component proportions.\\
In spite of our relatively low statistics compared with ARGOS, it is possible to compare the general characteristics of the two sets of MDFs in each field. Fields m1m10 and m10m8 show a reasonable agreement in the general shape compared with the stripes at $b=-10^{\circ}$ and $b=-7.5^{\circ}$ of \citet{argos_3}. For p1m4 and especially p0m6, our distributions seem to present a lack of B component compared with the $b=-5^{\circ}$ and $b=-7.5^{\circ}$ decompositions of ARGOS. The disagreement is even stronger for the p7m9 field, with a rather similar overall shape of the MDF, but displaced toward higher metallicity than in ARGOS. In general, our distributions follow the same vertical trend as ARGOS, changing the relative proportion of metallicity components with $b$, but  only roughly resembling the exact shapes given by the ARGOS fields.\\
To make the comparison to ARGOS more meaningful, we constructed a latitude strip from our data by co-adding the three fields p7m9, m1m10 and m10m8. The average latitude of this set is $-9\pm 1^{\circ}$. The GMM decomposition presents three metallicity components, the known (i) and (ii) from individual fields, and an additional component that adjusts the more defined metal poor tail bump. Comparing this with the $b=-10^{\circ}$ ARGOS strip still reveals the same differences.  We note that if we were to shift the ARGOS metallicity scale upwards by $\sim 0.2$~dex, the differences would be strongly alleviated: components A+B of ARGOS would fit  our component (i) nicely, while ARGOS component C would fit our (ii); furthermore, \citet{ness_splitted_rc} showed that stars with [Fe/H]$\geq -0.5$~dex display the X-shape, while we find this limit to be closer to $-0.2$ (see Sec.~\ref{sec:rc_split}), which again indicates a more metal-poor metallicity scale for ARGOS.\\

\begin{figure}
\begin{center}
\includegraphics[width=8.9cm]{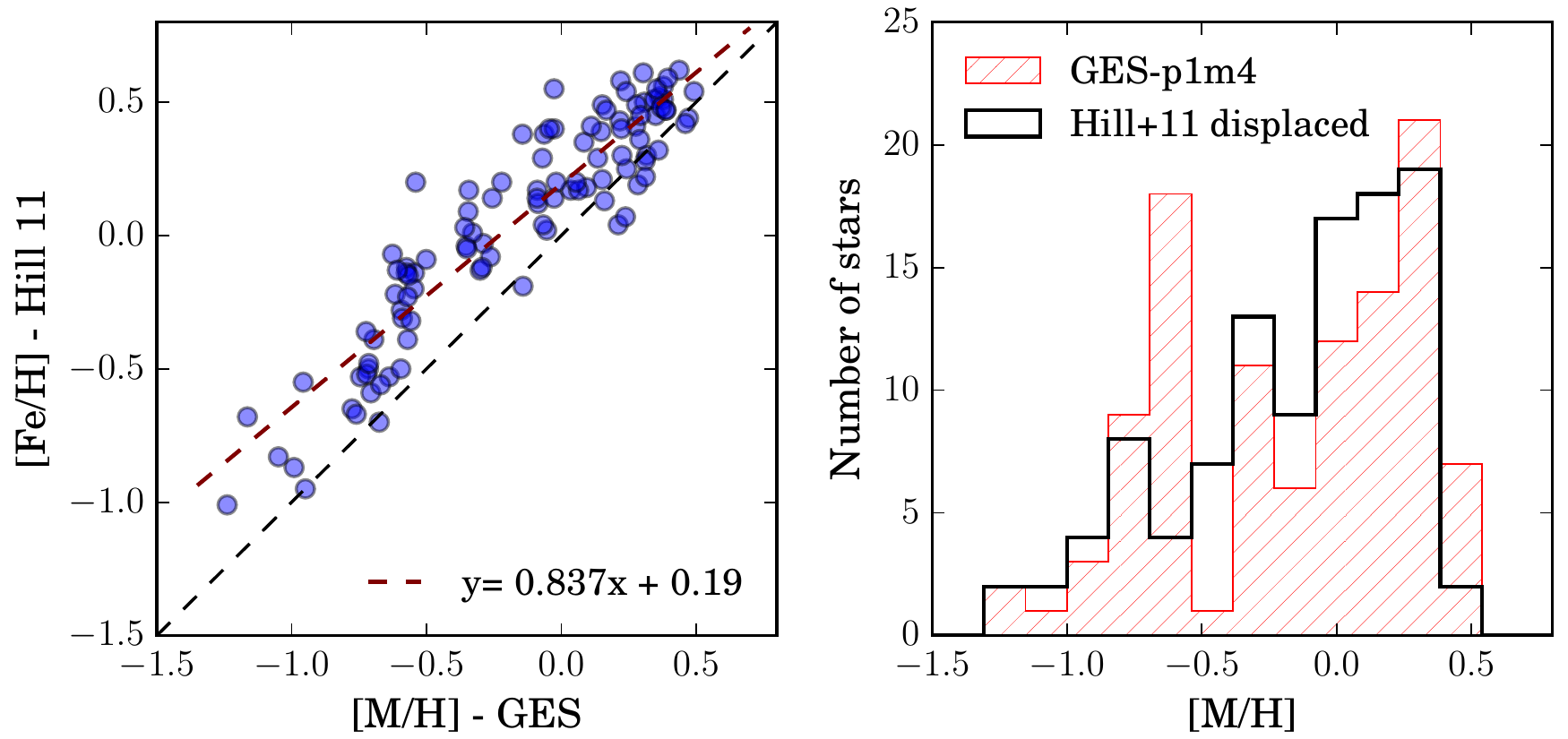}
\caption{Comparison of metallicity determinations for a set of common stars in between p1m4 sample and that studied in \citet{hill2011} in Baade's Window: \textit{left panel:} scatter plot of individual metallicity determinations.  \textit{b)}  the MDF from \citet{hill2011} has been displaced by 0.21 dex and is overploted with the MDF of the common stars from the field p1m4.}
\label{fig:comp_mdf_p1m4_vh}
  \end{center}
\end{figure}
In addition, we investigated the compatibility between the \citet{hill2011} MDF and the GES p1m4 MDF (same field). There are $\sim100$ stars in common because they were observed in the context of GES for calibration purposes (Sect. \ref{sec:datos}). From these stars, the comparison displayed in the left panel of Fig. \ref{fig:comp_mdf_p1m4_vh} was constructed.
This plot shows that there are both a shift and a weak trend between the metallicity scales. The shift can be quantified as the mean of the individual differences between the two sets of measurements. It corresponds to 0.21 dex.\\
We used a K-S test to investigate the compatibility between the respective MDFs. A \textit{p-value} of $0.013\%$ is obtained, which rejects the null hypothesis that both MDFs come from the same true metallicity distribution. Now, if we move the \citet{hill2011} distribution by $-0.21$~dex to account for the estimated shift, a K-S test gives a \textit{p-value} of $33\%$. Additionally, we shifted the MDF to try to simultaneously match the peaks of metallicity components. From this, we derived \textit{p-value} of $54\%$. It is clear that a metallicity zero-point adjustment is enough to make the two data sets comparable.\\
The color cut in the GES selection function is intended to be blue enough to sample the metal-poor tail of the  MDF (Sec. \ref{sec:datos}), while that of \citet{hill2011} was constructed to avoid the dwarf locus, which might heve biased the sample against the more metal-poor RGB stars. Despite the consequently possible small difference at the metal-poor tail of the MDF, the observed differences in the metallicity scale between the two estimations are relatively small and could be attributed to the different data (wavelength domain and resolution) and the methods of analysis employed to obtain the stellar parameters and metallicity.


\section{Double RC characterization}
\label{sec:rc_split}
\begin{figure}
  \begin{center}
\includegraphics[width=8.8cm]{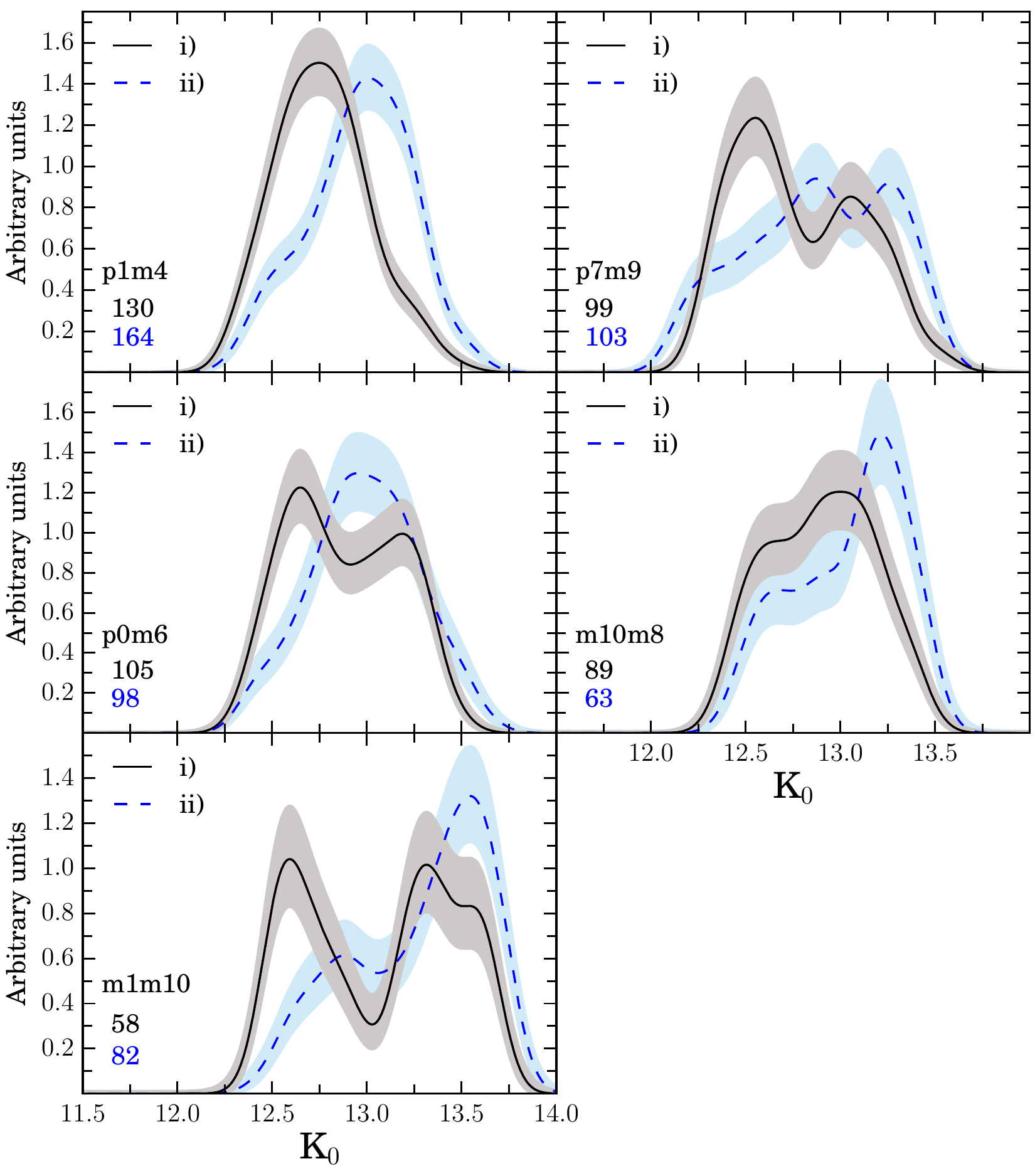}
\caption{Magnitude distributions for the metallicity components. The data are presented as generalized histograms computed with a kernel of $\sigma=0.1$ mag. We display the variability bands as shaded profiles around the curves.
Black (blue) solid (dashed) lines stand for the metal-rich (poor) MDF component.}
\label{fig:rc_split}
  \end{center}
\end{figure}

In their pioneer study, \citet{mcwilliam_zoccali2010} analyzed  luminosity functions of giant stars in the 2MASS survey. They found two distinct red clump populations  over a large bulge region ($\sim13^{\circ}$ in longitude and $20^{\circ}$ in latitude, with $|b|>5.5^{\circ}$). Then \citet{saito_x_shape} constructed a set of density maps from the same data that showed the structures traced by the RC clumps. The recent density maps of \citet{wegg_mapas}  from the  VVV DR1 K$_s$-band data demonstrate a single structure near the Galactic plane that splits into two components at high Galactic latitudes, a clear signature of an underlying X-shaped structure. \citet{ness_splitted_rc}  characterized the properties of the split RC structure and showed that the separation is visible  only for stars with \feh$>$-0.5 dex.\\
We  examine here the K$_s$ magnitude distributions in the GES fields along the minor and major axis. We first used the full VVV photometry in each field to construct luminosity functions of the giant branch (cf. Fig. \ref{fig:ejemplo_rc_cut}), in which we could trace double-clump structures. We verified that these clump luminosity substructures nicely trace the maps of \citet{wegg_mapas}, and then compared them with the distributions displayed by the spectroscopic subsamples.\\
Figure~\ref{fig:rc_split} shows the magnitude distributions of our clump stars in different metallicity ranges, corresponding to the components found in Sect. \ref{sec:distros_met}.
We built generalized histograms of K$_s$ magnitudes using a Gaussian kernel of $\sigma=0.1$ mag, and computed variability bands to assess the significance of the peaks revealed in the distribution. The photometric error of the VVV data at K$_s$=13 mag, is $\sim0.025$ mag in crowded bulge fields \citep{vvv_dr1}. Our larger kernel value is a compromise between the size of our sample, the size of the errors, and the magnitude separation we can expect to find between two RC components.

In every field were the general red clump luminosity function displays more than one structure (i.e. where the line of sight crosses more than one structure in the maps of  \citet{wegg_mapas}), the metal-rich component (i) also displays a bimodal K magnitude distribution. On the minor axis (left panel in Fig. \ref{fig:rc_split}), this occurs for $\rm |b|\leq 5$, so in our p0m6, and m1m10 fields, where the two peaks appear at K$_0=12.64$ and 13.18 mag, or K$_0=12.59$ and 13.32 mag, respectively. On the minor axis, p1m4 is too close to the plane for double red clump structures to appear because the arms of the X-shaped bulge, seen almost as an edge-on structure, merge in a central density of stars. Accordingly, the magnitude distribution of the metal-rich component (i) shows a single-peak. \\
To summarize, our metal-rich component (i) traces a gradual increase in the magnitude separation between the bright and faint peaks of the red clump luminosity histogram, corresponding to the arms of the X-shaped secular bulge structure. This reaches from an unresolved separation at p1m4 to  $\Delta$K$_0=0.54$ mag difference at p0m6, and 0.73 mag at m1m10. These numbers compare well with the separation of 0.65 mag found by \citet{mcwilliam_zoccali2010} for their field at $(l,b)=(1,-8)$.\\
Simultaneously, the metal-poor component (ii) in these fields along the minor axis shows only single-peak magnitude distributions. The mean magnitude for the metal-poor (ii) and the metal-rich (i) components are not always coincident, the metal-poor component being fainter on average. This is perhaps best seen in p1m4 where the magnitude distribution corresponding to the metal-rich component in our spectroscopic sample peaks  $0.26$ magnitudes brighter than the metal-poor component. A difference of $0.3$ mag is predicted by the isochrones between a metal-rich young RC and a metal-poor old RC \citep{bressan_isocronas} (e.g, -0.5 dex, 10 Gyr vs. 0.15 dex, 5 Gyr isochrones). A younger age for the metal-rich population than for the metal-poor population \citep[e.g,][]{bensby_edad_bulbo} could account for the observed shift between the luminosity functions of components (i) and (ii). This age effect may be accompanied (in particular far from the plane) by a volume effect whereby metal-poor stars are integrated to larger distances than the more centrally concentrated metal-rich stars.

The two off-axis fields (right-hand side of Fig. \ref{fig:rc_split}) also display interesting magnitude distributions. According to the maps of \citet{wegg_mapas} (mirrored in the red clump densities we built from the photometric samples), at latitudes of $-8$ to $-9^{\circ}$, we expect to cross only the near-side of the X-shape at positive longitudes (p7m9), and the far side of the X-shape at negative longitudes (m10m8). This latter line of sight is almost tangential to the bar before it hits the far side of the X. Indeed, the magnitude distribution of the metal-rich component (i) peaks at bright (resp. faint) magnitudes in p7m9 and m10m8. In the latter, the tail of brighter stars is still present in large proportion (outnumbering the metal-poor stars even at bright magnitudes), reflecting the fact that the line of sight grazes the bar in a large distance range.\\
These fields also present some peculiarities that are hard to understand and are perhaps related to those already noticed in their MDF (see Sect. \ref{sec:distros_met}). The line of sight at positive longitudes (p7m9) is expected to cross only the near side of the X-shape bulge \citep{wegg_mapas}. However, our spectroscopic sample spans a wide magnitude range, wider than expected from the X-shape itself, which would be traced by the main bright peak in the (i) metal-rich component in this field. In fact, there is a long tail of fainter stars, consisting of both metal-rich (i) and metal-poor (ii) stars. This tail, already present at low level in the maps of \citet{wegg_mapas}, would be expected to trace the smooth bulge component that we associate with the (ii) component; but we find it to contain a significant amount of metal-rich stars (i). This unexpected excess of metal-rich stars might well be related to the high mean metallicity in this  field.\\
 
At negative longitudes (m10m8), the magnitude distribution of component (ii) displays a unexpectedly strong peak at faint magnitudes. As the outermost field of our sample, the cone of sight integrates more and more stars with distance through the metal-poor extended component, which might be the reason for this excess.

In conclusion, it is clear that the double RC feature is  only seen for the metal-rich population, while metal-poor stars are distributed  more homogeneously across the magnitude range. Our data therefore suggest that the bulge contains at least two populations of different metallicity that belong to separated structures with distinct spatial distributions. When the line of sight intersects the two overdensities of the X-shaped bar, the apparent magnitude distribution of metal-rich stars display a bimodality that traces the bar structure, while metal-poor stars distribute homogeneously and traces some more extended and uniform structure.


\section{Kinematical features}
\label{sec:kinematical_features}
In this section, we characterize the substructures sampled by our RC-based metallicity components thanks to a detailed examination of the sample kinematics. For this purpose, we estimate the Galactocentric velocity V$_{\textmd{GC}}$ by correcting the radial heliocentric velocities V$_{\textmd{HC}}$ for the solar motion with respect to the Galactic center. 
We adopt V$_{rot}=220$ \kms as the magnitude of Galactic rotation at the distance of the Sun \citep{sol_rot_kerr}, and the velocity of the Sun with respect to the local standard of rest (LSR) to be 16.5 \kms in the direction ($l,b$)=($53^\circ,25^\circ$) \citep{mihalas_y_binney}.

\subsection{Rotation curve and velocity dispersion profiles}
\label{subsec:curvas_de_rot_y_disp}
\begin{figure}
\begin{center}
 \includegraphics[width=9cm]{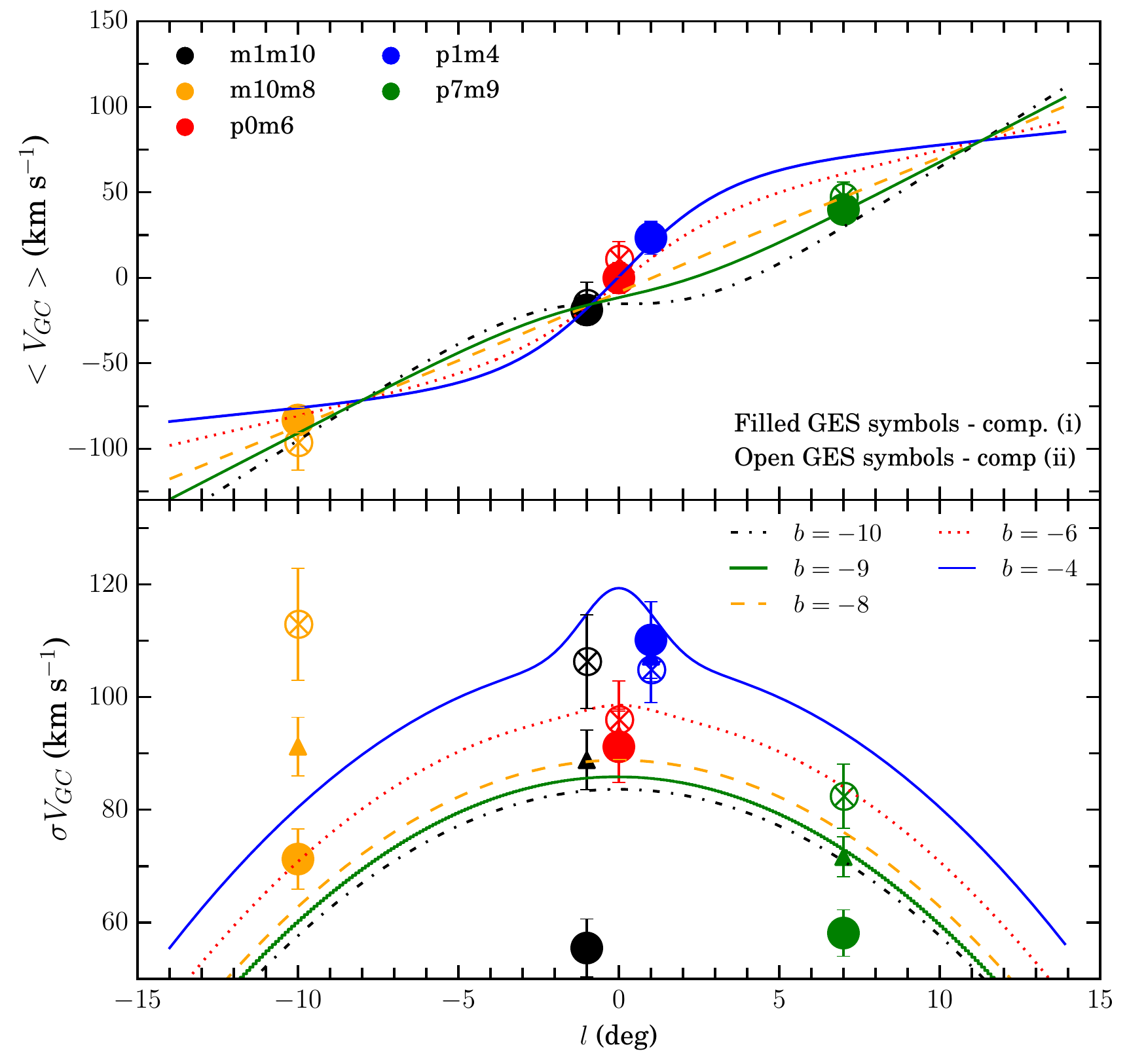}
  \caption{ Rotation curve (top) and velocity dispersion profile (bottom) for our fields. A set of curves computed from the GIBS kinematical maps are displayed using colors corresponding to the latitude of each field. To split stars into metallicity components, we adopt values tuned in each MDF, as presented in Sect. \ref{sec:rc_split}; \met$=-0.1$ dex for p0m6 and p7m9, -0.2 dex for m1m10, and -0.5 dex for m10m8. Filled and open symbols represent component (i) and (ii). Triangle colored symbols stand for the respective complete field samples.}
  \label{fig:curva_y_disp_de_vel}
  \end{center}
\end{figure}

We have used the computed Galactocentric velocities to construct the rotation and dispersion curves displayed in Fig. \ref{fig:curva_y_disp_de_vel}. For comparison purposes, we added profiles computed for every relevant latitude using equations (1) and (2) of \citet{gibs_paper} from the GIBS project. In their work, radial velocity and velocity dispersion maps of the Milky Way bulge are derived from $\sim6400$ RC stars in 24 fields spanning the area $-8^{\circ}<l<8^{\circ}$ and $-8^{\circ}<b<4^{\circ}$.

The \textit{s} shape of the rotation curve (upper panel of Fig. \ref{fig:curva_y_disp_de_vel}) reflects the projection of velocity vectors on the line of sight. The mean velocity values of our fields agree well with the more general curve traced by the GIBS maps. In particular, we verify with our data that the mean rotation velocity is independent of the height below the plane. This indicates a cylindrical rotation for the bulge. Moreover, in each field the two metallicity components present the same kinematical trend within the errors. This agrees with \citet{argos_4}.

On the other hand, the velocity dispersion in the different fields shows some striking differences with respect to the GIBS map. In our analysis, we introduced an independent piece of information,  using metallicity to split each sample into two components. When we compare the map (colored lines) with the triangle symbols, that is, using  all stars in each field, the agreement is reasonable, within errors, but some discrepancies are still visible in some fields. The most different results are seen in field m10m8, were the dispersion is underestimated in the GIBS map.\\
In general, our metallicity component analysis accounts for the divergent results. In fact, the velocity dispersion for the metal-poor component (ii) appears to be systematically higher, with values of about 100-110 \kms. On the other hand, the X-shaped bar metal-rich components (i) display similar or lower dispersions, changing with both $l$ and $b$.

\begin{figure}
\begin{center}
 \includegraphics[width=9cm]{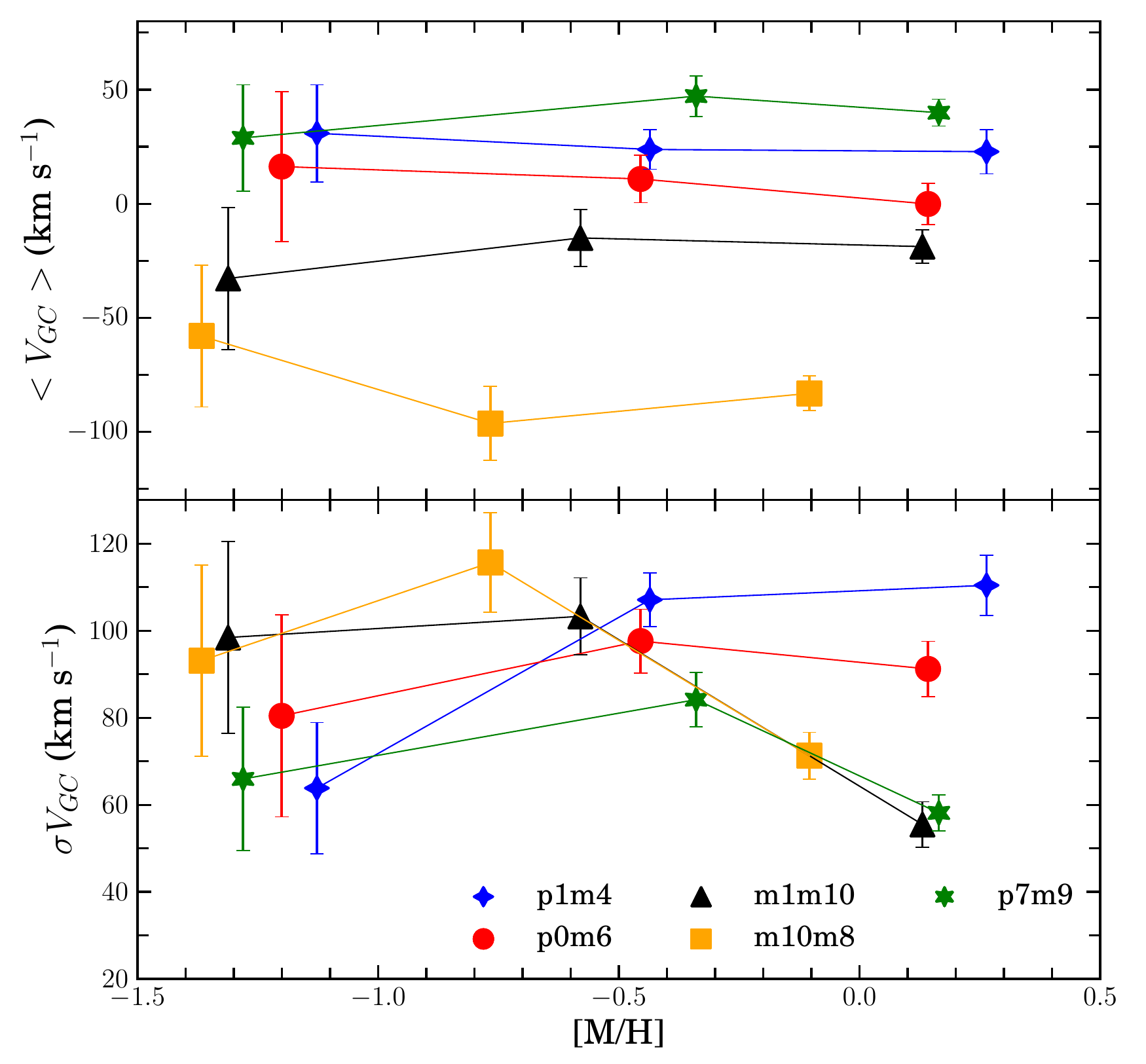}
  \caption{Mean and dispersion velocity versus metallicity. There are three points per field, corresponding to components (i), (ii),  and the metal-poor tail of stars with \met$<-1.0$ dex. The specific metallicity position of the points is given by the median value of the corresponding subsample.}
  \label{fig:curva_disp_mets}
  \end{center}
\end{figure}

\subsection{Metallicity dependence of the kinematics}
\label{subsec:kine_y_metalicidad_relaciones}

Figure \ref{fig:curva_disp_mets} provides a detailed view of the kinematics as a function of metallicity. The three points per field in each panel represent the two MDF components plus the metal-poor tail with \met$<-1.0$ dex.\\
The upper panel shows that the mean velocities have nearly flat profiles in all the fields. This suggests that all stars in a given line of sight, and thus all the structures to which they belong, rotate at the same rate. The specific field mean velocity values only depend on $l$ because of the different velocity vector projections on the line of sight. However, we note that with larger samples it is also possible to detect small variations with $b$ \citep{kunder_brava,argos_4,gibs_paper}.\\
The bottom panel of Fig.\ref{fig:curva_disp_mets} displays the velocity dispersion for the MDF components in each field. This figure shows that the differences in velocity dispersion between components (i) and (ii) change noticeably from field to field. Close to the midplane, both components have similar velocity dispersions, whereas far from the plane, the metal-rich component (i) has significantly lower velocity dispersions than (ii).

We first analyzed the kinematics of component (i). It becomes progressively cooler with increasing $|b|$ along the minor axis, from p1m4, p0m6 to m1m10. This change in the kinematics of metal-rich stars was previously noted and characterized in \citet{babusiaux2010}. The velocity dispersion of the positive-longitude field p7m9 component (i) is also very low, which agrees well with that of m1m10, located at a similar $b$. The two lines of sight intersect the near southern arm of the X-shaped bulge. At positive longitudes, the line of sight crosses a narrow region of the close arm, with orbits forming a relatively smooth stream, and producing low-velocity dispersions along the line of sight (although not necessarily in transverse motions). On the other hand, the metal-rich component (i) of m10m8 is kinematically hotter than its counterparts in other fields at similar latitudes. In this case,  that is, at negative longitudes, the line of sight samples the far southern arm of the X-shaped bulge. Because at the same latitude, the line of sight will intercept the arm farther from the plane and with a larger incidence angle at negative than at positive longitudes, we expect the physical size of the region sampled to be larger in the m10m8 direction. Furthermore, we might be reaching the tip of the arm, corresponding to the turnover point for most stellar orbits, where the speed and direction of the velocity vectors change. As a consequence, the line of sight integrates stars over a region where the radial projection of the velocity vectors of the stars is an inhomogeneous distribution, which explains the observed high-velocity dispersion.\\
The lower velocity dispersion at positive longitudes could also arise from a stronger contribution of inner Galactic disk foreground stars in the sample. This contamination is expected to be stronger in fields intersecting the near side of the bar because the RC luminosity functions of the bar and the disk blend more.\\

Meanwhile, the velocity dispersion of the metal-poor component (ii) along the minor axis (fields p1m4, p0m6, and m1m10) is high and constant within errors, around 100 \kms. The analysis of the two lateral fields shows similar results. Component (ii) of p7m9 has a somewhat smaller mean velocity dispersion, still in agreement within errors with those in minor axis fields. The velocity dispersion of component (ii) of m10m8, around 120\kms is the largest among the metal-poor components. We investigated whether this higher dispersion is related with the narrow peak at $\textmd{V}_{\textmd{rad}}\sim-140$ \kms of  the radial velocity distribution presented in Fig. \ref{fig:vrad_ges_bulbo}. The stars belonging to this peak correspond in equal size to the metal-rich and metal-poor components. Therefore, eliminating them from the sample does not decrease the dispersion significantly.\\ 
In conclusion, from this description, it is clear that the two metallicity populations identified in our fields display different kinematic signatures. The metal-poor spatially extended component (ii) appears to be systematically hot throughout the area sampled  by our fields. And in contrast, the kinematic properties of metal-rich component (i) are highly dependent on the specific position of the field, are generally hotter close to the plane and cooler with increasing ($l$,$b$). This depicts an inhomogeneous and more complex spatial orbital distribution for the structure traced by these metal-rich stars.\\
We can understand the observed kinematical properties of component (i) stars in terms of the orbital structure of the X-shaped bar that is traced by our metal-rich stars: an instability-driven mechanism that forms X-shaped structures starts from a disk instability  and takes stars from the internal disk(s) to form a bar. As a transient structure, the bar undergoes vertical instabilities, buckling and bending the stellar distribution, and resulting in a rearranged dynamical system consisting of many types of stellar orbit families. The global structure of this mix of orbits is complex (cf. Figs. 3 and 8 of \citeauthor{martinez_valpuesta_2006} \citeyear{martinez_valpuesta_2006}). One of the possible stable orbit families, sometimes named \textit{banana-like} orbits, has profiles that are up-bended or down-bended if seen edge-on, participating in an X-shaped structure. In this context, the observed variation in velocity dispersion for component (i) with $b$ might arise from the different mixture of orbit families on the line of sight. In the central parts of the bulge, the different families of orbits intersect, which builds a large velocity dispersion in these fields. In contrast, the line of sights intersect the arms of the X in external regions, where in general there are streams of dynamically similar stable banana orbits, which leads to lower velocity dispersions.


\subsection{Double RC kinematics and stream motions}
\label{subsec:double_rc_cin}
The RC magnitude distribution is double-peaked in the fields with the two arms of the X-shaped bulge. This is a signature of the metal-rich stars in component (i), as presented in Sect. \ref{sec:rc_split}. If the line of sight intersects both bulge overdensities, we can study the stream orbital motion of the stars by comparing their radial velocity in the two magnitude peaks \citep{sergio_x_kine}.
Table \ref{tab:split_velocidades_popI} gives detailed statistics for the split bright and faint groups in the two fields with double RC signature.\\
The p0m6 field corresponds to the location investigated by \citet{sergio_x_kine} with a larger sample of $\sim450$ stars. The mean velocities in the bright and faint RC overdensities reported in this work (V$_{\textmd{GC}}=-14.8\pm6.9$ and $9.22\pm6.7\  \textmd{km s}^{-1}$, respectively) agree with our measurements. The differences between our values and those of \citet{sergio_x_kine} arise because the values reported in Table \ref{tab:split_velocidades_popI} are only based  on the metal-rich stars of component (i)  (assumed to be those participating in the bar structure), while those of \citet{sergio_x_kine} are based on a sample built to populate only the two distinct peaks of the split RC, with no metallicity separation. This   
observational strategy, excluding the RC stars with intermediate luminosities between the two peaks, hampers their analysis of the metallicity-dependance of the RC bimodality. 
But a striking difference with our analysis arises on the metallicity dependence of the kinematics. 
Interestingly, \citet{sergio_x_kine} found that the velocity difference between the bright and the faint clumps is larger for stars with subsolar metallicity values, while our findings indicate the opposite, with the more metal-rich population (i) exhibiting this property as a characteristic imprint.

\begin{table} 
\centering
\caption{Statistics for bright and faint components of the metal-rich population (i) stars in fields with double RC.  Units in \kms.}
\begin{tabular}{llcc}
\hline
\hline
  &   & V$_{\textmd{GC}}$ bright & V$_{\textmd{GC}}$ faint \\\hline
p0m6  & Mean     & $-21.2\pm11.7$ & $24.8\pm12.9$   \\
      & $\sigma$ & $87.2\pm8.2$  & $89.6\pm9.2$       \\
      & Median   & $-17.1$        & $29.8$          \\
      & Number   &  57           &  48              \\\hline
m1m10 & Mean     & $-18.5\pm10.4$& $-19.2\pm10.5$   \\
      & $\sigma$ & $53.1\pm7.4$  & $57.5\pm7.4$     \\
      & Median   & $-13.8$       & $-23.6$          \\
      & Number   &  27           &  31              \\\hline
\end{tabular}
\label{tab:split_velocidades_popI}
\end{table}

The m1m10 field presents no radial velocity differences between the bright and faint populations, with equal mean, median, and dispersion values, within errors. This result is consistent with that presented in \citet{uttenthaler_2012} from a red giant sample of 401 stars at ($l$,$b$)=(0,-10). They did not find a statistically significant difference in velocity for the bright and faint groups, both with a similar mean value of $\sim-10$~\kms. As an additional comparison, \citet{de_propris_2011} found no difference between the clump velocity distributions for a field at $(l,b)=(0^\circ,-8^\circ)$ from their-low resolution sample of $\sim700$ stars.

On the other hand, several studies \citep{rangwala_2009,babusiaux2010,babusiaux2014} have detected stream motions in internal bulge fields where the red clump has a single-peaked magnitude distribution. In these cases, the line of sight still intersects complex family orbits and can be used to probe their kinematical structure. Therefore, we searched for stream motions in the  p1m4 field by separating stars brighter and fainter than the peak of the magnitude distribution of component (i). This enabled us to probe for differences in kinematics between stars located at either sides of the Galactic bar. The mean velocity is $\textmd{V}_{\textmd{GC}}=-7.1\pm13.0$~\kms for the bright group and $52.2\pm13.1$~\kms for the faint group, with the same velocity dispersion. A K-S test for the two velocity distributions results in a statistically significant difference, with a p-value of $0.12\%$. In addition, the mean velocity difference is diluted ($19.3\pm8.2$~\kms for the bright, and $31.4\pm8.9$~\kms for the faint group) when the stars of component (ii) are included in the analysis. This agrees with \citet{babusiaux2010}, who also detected the stream motions only for metal-rich stars, selecting those with \feh>-0.09 dex.

In conclusion, we reassert once again that the X-shaped bar is seen only in the metal-rich component (i), seen this time from the presence of streaming motions: wherever they are detected, streaming motions are confined to the metal-rich (i) component.
The orbital structure in the metal-rich X-shaped bar is complex because it consists of the superposition of several stable family orbits. The radial velocity allows us to trace the homogeneity on a given line of sight. In this way, the different results for different fields reveal the nature of the orbit streams in the branches of the X-shape bulge. 


\section{Discussion and conclusions}
\label{sec:conclusiones}
We have made use of data from the internal Data Release 1 of the Gaia-ESO survey to analyze a sample of $\sim1200$ stars in five fields in the bulge region located at $-10<l<7$ and $-10<b<-4$. Using VISTA photometry, we verified the consistency of the stellar atmospheric parameters by splitting giants from disk dwarfs. From this giant clean sample, we determined extinction values in each line of sight that compare well with the recent determinations of \citet{oscarMapas_extincion}. We used this as an additional sanity check for the internal consistence of the atmospheric parameters, which as a by-product yielded extinction values in each field.

Then, we photometrically selected a subsample of bona fide bulge RC stars. Some disk RGB contaminants remains in this selection because they are in the same magnitude region. Based on this subsample, we investigated the MDF substructure and its relation with kinematics. 

Using a GMM decomposition, we found two metallicity components that make up the global MDF shape. This general MDF bimodality consist of a metal-rich narrow component (i) at +0.10 to +0.20 dex and a broader one (ii) at subsolar values. We note that other studies, using larger samples, argued for a more complex substructure \citep{argos_3}. For our sample, the two components suggest that the stars in the bulge constitute a composite population that overlaps on the line of sight. A vertical gradient is also found through the minor axis, whose nature can be explained following the interpretation of \citet{zoccali2008}. In fact, the proportion of stars belonging to both metallicity components changes with increasing latitude, with a progressively larger metal-poor population in detriment of the metal-rich population, which diminishes with $b$. This variation in the component mixture changes the shape of the MDF and in turn the global mean metallicity, reproducing the observed metallicity gradient. If we extrapolate this trend to the central bulge regions, we could expect a clear predominance of stars belonging to the metal-rich component, which makes the metallicity gradient constant to the first order, as claimed in some studies for $b<-4$ \citep{rich_no_gradiente, Rich_2012_no_gradiente}.

We then used kinematical and photometrical data to further characterize the metallicity components. This analysis led us to propose that there are two structurally different populations in the line of sight that correspond to the two  MDF components.

The metal-rich component (i), with supersolar metallicity, presents small mean metallicity variations between the fields. It is kinematically colder than the metal-poor component (ii), but with a variable $\sigma V_{\textmd{GC}}$  with ($l$,$b$). In some fields, this metal-rich population has a double-peaked magnitude distribution, consistent with an observed bimodality in the general field luminosity function. This population was interpreted as a pseudobulge structure formed over a long time-scale through secular disk evolution. The initial instability drives a reorganization of the stars, producing a stable orbital structure with some of the family orbits forming an X-shaped structure. A metallicity gradient is predicted in some models for this structure, in which the mixing during the bar and buckling instabilities is incomplete. Therefore, any preexisting radial gradient in the initial disk is transferred to some degree to the newly formed structure as a vertical gradient \citep{MV_gradiente}.

The second, metal-poor component (ii), with subsolar metallicity, presents a clear negative metallicity gradient with increasing ($l$,$b$). The $\sigma V_{\textmd{GC}}$ values present small variations from field to field, indicating a uniform kinematically hot structure. The respective field magnitude distributions are flat or single peaked, without statistically strong signatures of double RC. We interpreted this component as an extended structure, possibly a classical spheroid, formed on a rapid time-scale. It coexists with the bar, surrounding it in the bulge region.  From a sample of $7663$ RR-Lyrae stars that sampled the old metal-poor population, \citet{dekany2013} found that their distribution in the bulge shows no X-shape and not even a bar, but has a more spheroidal centrally concentrated distribution. As a dissipatively collapsed structure, this component would be expected to present a radial metallicity gradient \citep{larson1974}, with the more metal-poor stars in the outskirts, similarly to what is observed. 

In this scenario, a line of sight at a given ($l$,$b$) crosses a fraction of stars that belong to the metal-rich boxy bar (if visible) plus a group of spheroid stars that become more metal-poor with increasing distances from the bulge center. This scenario is in conflict with the N-body Galaxy model of \citet{shen2010}, who constrained any classical bulge contribution to be lower than $8\%$ of the disk mass. The model of \citet{shen2010} fits the BRAVA data nicely when a pure disk galaxy is modeled. But the BRAVA data  have no metallicity information. Because the BRAVA data are for M giants, they might even be biased toward metal-rich stars. We showed that the kinematics behaves differently for the metal-rich and metal-poor MDF components, which means that the characteristics of the global distribution depend on the proportion of stars of each component in a given field. In the same line, \citet{babusiaux2010} found a variation of kinematics with metallicity on the minor axis, with the $\sigma V_{\textmd{r}}$ behavior of metal-rich and metal-poor stars being reproduced by the disk and spheroid populations of the model of \citet{fux_barra}. Furthermore, the coexistence of classically and seculary evolved pseudobulge structures has been observed in external galaxies \citep{Peletier_2007,Erwin_2008} and is also predicted by N-body simulations \citep{athanassoula_2005}. It is interesting to note that while the two metallicity components have different $\sigma V_{\textmd{GC}}$ profiles, they behave similarly in terms of mean rotational velocity (Fig. \ref{fig:curva_y_disp_de_vel}). This is not necessarily expected if only component (i) participates in the X-shaped bar/bulge, but it might be explained as a dynamical influence of this structure on the spheroid kinematics \citep{saha_m_valpuesta,bekki2011}. 

Although it is clearly beyond the scope of the present paper to estimate which mass fraction of the bulge would be in this spheroid population (insufficient spatial coverage as yet in GES, not large enough samples),  it is interesting to note that the proportion of the metal-poor (ii) component in all of our fields is significant. This might be interpreted, with the caveat of our limited statistics and spatial coverage, as an indication for a structure that probably has a similar mass-content as the bar structure. We also note that the relationship of this spheroid population with the very inner parts of the Galactic thick disk is only poorly established (chemical parenthood of the two populations, see, for example, \citeauthor{melendez_similaridad} \citeyear{melendez_similaridad}, \citeauthor{hill2011} \citeyear{hill2011}, \citeauthor{oscar_similaridad} \citeyear{oscar_similaridad}, \citeauthor{bensby_edad_bulbo} \citeyear{bensby_edad_bulbo}), which leaves the possibility open that part of this rotating spheroidal structure might be part of the inner thick disc.\\
This scenario has recently been examined by \citet{di_matteo2014}, who argued that the characteristics (kinematics, chemistry) of the metal-poor bulge component agree better with a thick-disk nature than with a spheroid. In particular, they argued that a classical bulge spheroid with the mean metallicity of ARGOS C cannot be larger than $10^8\textmd{M}_{\odot}$ \citep{gallazzi2005}, which is  incompatible  with the number of stars in that component.\\
If we assume a thick-disk origin for the metal-poor component, the similar V$_{rot}$ of the (ii) and (i) components can be understood because both discs are affected by the bar potential \citep{bekki2011}.
However, our component (ii) is slightly more metal-rich than component C of ARGOS and would allow for a $5\times10^{9}\textmd{M}_{\odot}$, or even up to $10^{10}\textmd{M}_{\odot}$, this time compatible with a metal-poor component accounting for $\sim50\%$ of the total bulge mass. Furthermore, we pointed out in Sect. \ref{subsec:gradiente_vertical} that our metal-poor component displays a vertical gradient of its own, and although current models are not really designed to address this particular question, if the initial thick disk had no radial gradient \citep[][accepted]{recio-blanco2014,mikolaitis2014}, no vertical gradient in the final inner thick disk is expected \citep[][ Fig. 5]{bekki2011}.

Finally, one of the key objectives of Gaia-ESO survey is to provide the first homogeneous overview of the distributions of kinematics and elemental abundances for all main components of the Milky Way. In particular, the bulge has just recently begun to be explored using multiple fields to probe the variation of its properties. The sample analyzed here shows that an approach that combines metallicity, kinematics, and an appropriate field sampling allows distinguishing the different bulge populations. It also stresses the relevance and necessity of theoretical models that combine kinematic and chemical evolution to interpret the increasing amount of observational data.

\begin{acknowledgements}
A. Recio-Blanco, P. de Laverny and V. Hill acknowledge the support of the French Agence Nationale de la Recherche, under contract ANR-2010-BLAN-0508-01OTP, and the Programme National de Cosmologie et Galaxies. This work was partly supported by the European Union FP7 programme through ERC grant number 320360 and by the Leverhulme Trust through grant RPG-2012-541. We acknowledge the support from INAF and Ministero dell' Istruzione, dell' Universit\`a' e della Ricerca (MIUR) in the form of the grant "Premiale VLT 2012". The results presented here benefit from discussions held during the Gaia-ESO workshops and conferences supported by the ESF (European Science Foundation) through the GREAT Research Network Programme. MZ acknowledge support by Proyecto Fondecyt Regular 1110393, by the BASAL Center for Astrophysics and Associated Technologies PFB-06, and by the Chilean Ministry for the Economy, Development, and Tourism’s 
Programa Iniciativa Científica Milenio through grant IC120009 awarded to the "Millennium Institute of Astrophysics (MAS)". A.Rojas-Arriagada thanks the organizers and participants of the conference ``Formation and Evolution of the Galactic Bulge'', Sexten, Italy, January 2014, for the stimulating and productive discussion atmosphere. T.B. was funded by grant No. 621-2009-3911 from The Swedish Research Council. 
\end{acknowledgements}


 \bibliographystyle{aa}
\bibliography{bulbo_ges} 

\begin{thebibliography}{71}
\expandafter\ifx\csname natexlab\endcsname\relax\def\natexlab#1{#1}\fi

\bibitem[{{Athanassoula}(2003)}]{athanassoula2003}
{Athanassoula}, E. 2003, \mnras, 341, 1179

\bibitem[{{Athanassoula}(2005)}]{athanassoula_2005}
{Athanassoula}, E. 2005, \mnras, 358, 1477

\bibitem[{{Babusiaux} {et~al.}(2010){Babusiaux}, {G{\'o}mez}, {Hill}, {Royer},
  {Zoccali}, {Arenou}, {Fux}, {Lecureur}, {Schultheis}, {Barbuy}, {Minniti}, \&
  {Ortolani}}]{babusiaux2010}
{Babusiaux}, C., {G{\'o}mez}, A., {Hill}, V., {et~al.} 2010, \aap, 519, A77

\bibitem[{{Babusiaux} {et~al.}(2014){Babusiaux}, {Katz}, {Hill}, {Royer},
  {Gomez}, {Arenou}, {Combes}, {Di Matteo}, {Gilmore}, {Haywood}, {Robin},
  {Rodriguez-Fernandez}, {Sartoretti}, \& {Schultheis}}]{babusiaux2014}
{Babusiaux}, C., {Katz}, D., {Hill}, V., {et~al.} 2014, ArXiv e-prints

\bibitem[{{Bekki} \& {Tsujimoto}(2011)}]{bekki2011}
{Bekki}, K. \& {Tsujimoto}, T. 2011, \mnras, 416, L60

\bibitem[{{Bensby} {et~al.}(2011){Bensby}, {Ad{\'e}n}, {Mel{\'e}ndez}, {Gould},
  {Feltzing}, {Asplund}, {Johnson}, {Lucatello}, {Yee}, {Ram{\'{\i}}rez},
  {Cohen}, {Thompson}, {Bond}, {Gal-Yam}, {Han}, {Sumi}, {Suzuki}, {Wada},
  {Miyake}, {Furusawa}, {Ohmori}, {Saito}, {Tristram}, \&
  {Bennett}}]{bensby2011}
{Bensby}, T., {Ad{\'e}n}, D., {Mel{\'e}ndez}, J., {et~al.} 2011, \aap, 533,
  A134

\bibitem[{{Bensby} {et~al.}(2013){Bensby}, {Yee}, {Feltzing}, {Johnson},
  {Gould}, {Cohen}, {Asplund}, {Mel{\'e}ndez}, {Lucatello}, {Han}, {Thompson},
  {Gal-Yam}, {Udalski}, {Bennett}, {Bond}, {Kohei}, {Sumi}, {Suzuki}, {Suzuki},
  {Takino}, {Tristram}, {Yamai}, \& {Yonehara}}]{bensby_edad_bulbo}
{Bensby}, T., {Yee}, J.~C., {Feltzing}, S., {et~al.} 2013, \aap, 549, A147

\bibitem[{{Bressan} {et~al.}(2012){Bressan}, {Marigo}, {Girardi}, {Salasnich},
  {Dal Cero}, {Rubele}, \& {Nanni}}]{bressan_isocronas}
{Bressan}, A., {Marigo}, P., {Girardi}, L., {et~al.} 2012, \mnras, 427, 127

\bibitem[{{Bureau} {et~al.}(2006){Bureau}, {Aronica}, {Athanassoula},
  {Dettmar}, {Bosma}, \& {Freeman}}]{bureau_2006}
{Bureau}, M., {Aronica}, G., {Athanassoula}, E., {et~al.} 2006, \mnras, 370,
  753

\bibitem[{{de Laverny} {et~al.}(2012){de Laverny}, {Recio-Blanco}, {Worley}, \&
  {Plez}}]{grid_ambre}
{de Laverny}, P., {Recio-Blanco}, A., {Worley}, C.~C., \& {Plez}, B. 2012,
  \aap, 544, A126

\bibitem[{{De Propris} {et~al.}(2011){De Propris}, {Rich}, {Kunder}, {Johnson},
  {Koch}, {Brough}, {Conselice}, {Gunawardhana}, {Palamara}, {Pimbblet}, \&
  {Wijesinghe}}]{de_propris_2011}
{De Propris}, R., {Rich}, R.~M., {Kunder}, A., {et~al.} 2011, \apjl, 732, L36

\bibitem[{{Debattista} {et~al.}(2006){Debattista}, {Mayer}, {Carollo}, {Moore},
  {Wadsley}, \& {Quinn}}]{debattista_2006}
{Debattista}, V.~P., {Mayer}, L., {Carollo}, C.~M., {et~al.} 2006, \apj, 645,
  209

\bibitem[{{D{\'e}k{\'a}ny} {et~al.}(2013){D{\'e}k{\'a}ny}, {Minniti},
  {Catelan}, {Zoccali}, {Saito}, {Hempel}, \& {Gonzalez}}]{dekany2013}
{D{\'e}k{\'a}ny}, I., {Minniti}, D., {Catelan}, M., {et~al.} 2013, \apjl, 776,
  L19

\bibitem[{{Di Matteo} {et~al.}(2014){Di Matteo}, {Haywood}, {Gomez}, {van
  Damme}, {Combes}, {Halle}, {Semelin}, {Lehnert}, \& {Katz}}]{di_matteo2014}
{Di Matteo}, P., {Haywood}, M., {Gomez}, A., {et~al.} 2014, ArXiv e-prints

\bibitem[{{Dwek} {et~al.}(1995){Dwek}, {Arendt}, {Hauser}, {Kelsall}, {Lisse},
  {Moseley}, {Silverberg}, {Sodroski}, \& {Weiland}}]{dwek}
{Dwek}, E., {Arendt}, R.~G., {Hauser}, M.~G., {et~al.} 1995, \apj, 445, 716

\bibitem[{{Eggen} {et~al.}(1962){Eggen}, {Lynden-Bell}, \&
  {Sandage}}]{eggen_1962}
{Eggen}, O.~J., {Lynden-Bell}, D., \& {Sandage}, A.~R. 1962, \apj, 136, 748

\bibitem[{{Erwin}(2008)}]{Erwin_2008}
{Erwin}, P. 2008, in IAU Symposium, Vol. 245, IAU Symposium, ed. M.~{Bureau},
  E.~{Athanassoula}, \& B.~{Barbuy}, 113--116

\bibitem[{{Falc{\'o}n-Barroso} {et~al.}(2004){Falc{\'o}n-Barroso}, {Bacon},
  {Bureau}, {Cappellari}, {Davies}, {Emsellem}, {Krajnovi{\'c}}, {Kuntschner},
  {McDermid}, {Peletier}, \& {de Zeeuw}}]{falcon_barroso}
{Falc{\'o}n-Barroso}, J., {Bacon}, R., {Bureau}, M., {et~al.} 2004,
  Astronomische Nachrichten, 325, 92

\bibitem[{{Fukugita} {et~al.}(1998){Fukugita}, {Hogan}, \&
  {Peebles}}]{fukugita}
{Fukugita}, M., {Hogan}, C.~J., \& {Peebles}, P.~J.~E. 1998, \apj, 503, 518

\bibitem[{{Fux}(1999)}]{fux_barra}
{Fux}, R. 1999, \aap, 345, 787

\bibitem[{{Gallazzi} {et~al.}(2005){Gallazzi}, {Charlot}, {Brinchmann},
  {White}, \& {Tremonti}}]{gallazzi2005}
{Gallazzi}, A., {Charlot}, S., {Brinchmann}, J., {White}, S.~D.~M., \&
  {Tremonti}, C.~A. 2005, \mnras, 362, 41

\bibitem[{{Gilmore} {et~al.}(2012){Gilmore}, {Randich}, {Asplund}, {Binney},
  {Bonifacio}, {Drew}, {Feltzing}, {Ferguson}, {Jeffries}, {Micela},
  {Negueruela}, {Prusti}, {Rix}, {Vallenari}, {Alfaro}, {Allende-Prieto},
  {Babusiaux}, {Bensby}, {Blomme}, {Bragaglia}, {Flaccomio}, {Fran{\c c}ois},
  {Irwin}, {Koposov}, {Korn}, {Lanzafame}, {Pancino}, {Paunzen},
  {Recio-Blanco}, {Sacco}, {Smiljanic}, {Van Eck}, \& {Walton}}]{GESMessenger}
{Gilmore}, G., {Randich}, S., {Asplund}, M., {et~al.} 2012, The Messenger, 147,
  25

\bibitem[{{Gonzalez} {et~al.}(2011{\natexlab{a}}){Gonzalez}, {Rejkuba},
  {Zoccali}, {Hill}, {Battaglia}, {Babusiaux}, {Minniti}, {Barbuy},
  {Alves-Brito}, {Renzini}, {Gomez}, \& {Ortolani}}]{oscar_similaridad}
{Gonzalez}, O.~A., {Rejkuba}, M., {Zoccali}, M., {et~al.} 2011{\natexlab{a}},
  \aap, 530, A54

\bibitem[{{Gonzalez} {et~al.}(2013){Gonzalez}, {Rejkuba}, {Zoccali}, {Valent},
  {Minniti}, \& {Tobar}}]{gonzalez_metallicity_map}
{Gonzalez}, O.~A., {Rejkuba}, M., {Zoccali}, M., {et~al.} 2013, \aap, 552, A110

\bibitem[{{Gonzalez} {et~al.}(2011{\natexlab{b}}){Gonzalez}, {Rejkuba},
  {Zoccali}, {Valenti}, \& {Minniti}}]{oscarMapas_extincion}
{Gonzalez}, O.~A., {Rejkuba}, M., {Zoccali}, M., {Valenti}, E., \& {Minniti},
  D. 2011{\natexlab{b}}, \aap, 534, A3

\bibitem[{{Gonz{\'a}lez Hern{\'a}ndez} \& {Bonifacio}(2009)}]{calib_bonifacio}
{Gonz{\'a}lez Hern{\'a}ndez}, J.~I. \& {Bonifacio}, P. 2009, \aap, 497, 497

\bibitem[{{Harris}(1996)}]{harris}
{Harris}, W.~E. 1996, \aj, 112, 1487

\bibitem[{{Hill} {et~al.}(2011){Hill}, {Lecureur}, {G{\'o}mez}, {Zoccali},
  {Schultheis}, {Babusiaux}, {Royer}, {Barbuy}, {Arenou}, {Minniti}, \&
  {Ortolani}}]{hill2011}
{Hill}, V., {Lecureur}, A., {G{\'o}mez}, A., {et~al.} 2011, \aap, 534, A80

\bibitem[{{Howard} {et~al.}(2008){Howard}, {Rich}, {Reitzel}, {Koch}, {De
  Propris}, \& {Zhao}}]{Howard_brava_selection}
{Howard}, C.~D., {Rich}, R.~M., {Reitzel}, D.~B., {et~al.} 2008, \apj, 688,
  1060

\bibitem[{{Immeli} {et~al.}(2004){Immeli}, {Samland}, {Gerhard}, \&
  {Westera}}]{immeli_2004}
{Immeli}, A., {Samland}, M., {Gerhard}, O., \& {Westera}, P. 2004, \aap, 413,
  547

\bibitem[{{Jofre} {et~al.}(2013){Jofre}, {Heiter}, {Soubiran},
  {Blanco-Cuaresma}, {Worley}, {Pancino}, {Bergemann}, {Cantat-Gaudin},
  {Gonzalez-Hernandez}, {Hill}, {Lardo}, {de Laverny}, {Lind}, {Magrini},
  {Masseron}, {Montes}, {Mucciarelli}, {Nordlander}, {Recio-Blanco}, {Sobeck},
  {Sordo}, {Sousa}, {Tabernero}, {Vallenari}, \& {Van
  Eck}}]{estrellas_benchmark}
{Jofre}, P., {Heiter}, U., {Soubiran}, C., {et~al.} 2013, ArXiv e-prints

\bibitem[{{Kerr} \& {Lynden-Bell}(1986)}]{sol_rot_kerr}
{Kerr}, F.~J. \& {Lynden-Bell}, D. 1986, Highlights of Astronomy, 7, 889

\bibitem[{{Kordopatis} {et~al.}(2011){Kordopatis}, {Recio-Blanco}, {de
  Laverny}, {Bijaoui}, {Hill}, {Gilmore}, {Wyse}, \&
  {Ordenovic}}]{kordopatis_matisse}
{Kordopatis}, G., {Recio-Blanco}, A., {de Laverny}, P., {et~al.} 2011, \aap,
  535, A106

\bibitem[{{Kormendy} \& {Kennicutt}(2004)}]{kormendy_kennicutt_review}
{Kormendy}, J. \& {Kennicutt}, Jr., R.~C. 2004, \araa, 42, 603

\bibitem[{{Kunder} {et~al.}(2012){Kunder}, {Koch}, {Rich}, {de Propris},
  {Howard}, {Stubbs}, {Johnson}, {Shen}, {Wang}, {Robin}, {Kormendy}, {Soto},
  {Frinchaboy}, {Reitzel}, {Zhao}, \& {Origlia}}]{kunder_brava}
{Kunder}, A., {Koch}, A., {Rich}, R.~M., {et~al.} 2012, \aj, 143, 57

\bibitem[{{Larson}(1974)}]{larson1974}
{Larson}, R.~B. 1974, \mnras, 166, 585

\bibitem[{{L{\'o}pez-Corredoira} {et~al.}(2005){L{\'o}pez-Corredoira},
  {Cabrera-Lavers}, \& {Gerhard}}]{lopez_corredoira_boxy}
{L{\'o}pez-Corredoira}, M., {Cabrera-Lavers}, A., \& {Gerhard}, O.~E. 2005,
  \aap, 439, 107

\bibitem[{{Martinez-Valpuesta} \& {Gerhard}(2013)}]{MV_gradiente}
{Martinez-Valpuesta}, I. \& {Gerhard}, O. 2013, \apjl, 766, L3

\bibitem[{{Martinez-Valpuesta} {et~al.}(2006){Martinez-Valpuesta}, {Shlosman},
  \& {Heller}}]{martinez_valpuesta_2006}
{Martinez-Valpuesta}, I., {Shlosman}, I., \& {Heller}, C. 2006, \apj, 637, 214

\bibitem[{{McWilliam} \& {Zoccali}(2010)}]{mcwilliam_zoccali2010}
{McWilliam}, A. \& {Zoccali}, M. 2010, \apj, 724, 1491

\bibitem[{{Mel{\'e}ndez} {et~al.}(2008){Mel{\'e}ndez}, {Asplund},
  {Alves-Brito}, {Cunha}, {Barbuy}, {Bessell}, {Chiappini}, {Freeman},
  {Ram{\'{\i}}rez}, {Smith}, \& {Yong}}]{melendez_similaridad}
{Mel{\'e}ndez}, J., {Asplund}, M., {Alves-Brito}, A., {et~al.} 2008, \aap, 484,
  L21

\bibitem[{{Mihalas} \& {Binney}(1981)}]{mihalas_y_binney}
{Mihalas}, D. \& {Binney}, J. 1981, {Galactic astronomy: Structure and
  kinematics /2nd edition/}

\bibitem[{{Mikolaitis} {et~al.}(2014){Mikolaitis}, {Hill}, {Recio-Blanco}, \&
  et~al.}]{mikolaitis2014}
{Mikolaitis}, S., {Hill}, V., {Recio-Blanco}, A., \& et~al. 2014, \aap

\bibitem[{{Minniti} {et~al.}(2010){Minniti}, {Lucas}, {Emerson}, {Saito},
  {Hempel}, {Pietrukowicz}, {Ahumada}, {Alonso}, {Alonso-Garcia}, {Arias},
  {Bandyopadhyay}, {Barb{\'a}}, {Barbuy}, {Bedin}, {Bica}, {Borissova},
  {Bronfman}, {Carraro}, {Catelan}, {Clari{\'a}}, {Cross}, {de Grijs},
  {D{\'e}k{\'a}ny}, {Drew}, {Fari{\~n}a}, {Feinstein}, {Laj{\'u}s}, {Gamen},
  {Geisler}, {Gieren}, {Goldman}, {Gonzalez}, {Gunthardt}, {Gurovich},
  {Hambly}, {Irwin}, {Ivanov}, {Jord{\'a}n}, {Kerins}, {Kinemuchi}, {Kurtev},
  {L{\'o}pez-Corredoira}, {Maccarone}, {Masetti}, {Merlo}, {Messineo},
  {Mirabel}, {Monaco}, {Morelli}, {Padilla}, {Palma}, {Parisi}, {Pignata},
  {Rejkuba}, {Roman-Lopes}, {Sale}, {Schreiber}, {Schr{\"o}der}, {Smith}, {},
  {Soto}, {Tamura}, {Tappert}, {Thompson}, {Toledo}, {Zoccali}, \&
  {Pietrzynski}}]{Minniti_vvv}
{Minniti}, D., {Lucas}, P.~W., {Emerson}, J.~P., {et~al.} 2010, \na, 15, 433

\bibitem[{{Nataf} {et~al.}(2010){Nataf}, {Udalski}, {Gould}, {Fouqu{\'e}}, \&
  {Stanek}}]{nataf2010}
{Nataf}, D.~M., {Udalski}, A., {Gould}, A., {Fouqu{\'e}}, P., \& {Stanek},
  K.~Z. 2010, \apjl, 721, L28

\bibitem[{{Ness} {et~al.}(2013{\natexlab{a}}){Ness}, {Freeman}, {Athanassoula},
  {Wylie-de-Boer}, {Bland-Hawthorn}, {Asplund}, {Lewis}, {Yong}, {Lane}, \&
  {Kiss}}]{argos_3}
{Ness}, M., {Freeman}, K., {Athanassoula}, E., {et~al.} 2013{\natexlab{a}},
  \mnras, 430, 836

\bibitem[{{Ness} {et~al.}(2013{\natexlab{b}}){Ness}, {Freeman}, {Athanassoula},
  {Wylie-de-Boer}, {Bland-Hawthorn}, {Asplund}, {Lewis}, {Yong}, {Lane},
  {Kiss}, \& {Ibata}}]{argos_4}
{Ness}, M., {Freeman}, K., {Athanassoula}, E., {et~al.} 2013{\natexlab{b}},
  \mnras, 432, 2092

\bibitem[{{Ness} {et~al.}(2012){Ness}, {Freeman}, {Athanassoula},
  {Wylie-De-Boer}, {Bland-Hawthorn}, {Lewis}, {Yong}, {Asplund}, {Lane},
  {Kiss}, \& {Ibata}}]{ness_splitted_rc}
{Ness}, M., {Freeman}, K., {Athanassoula}, E., {et~al.} 2012, \apj, 756, 22

\bibitem[{{Peletier} {et~al.}(2007){Peletier}, {Falc{\'o}n-Barroso}, {Bacon},
  {Cappellari}, {Davies}, {de Zeeuw}, {Emsellem}, {Ganda}, {Krajnovi{\'c}},
  {Kuntschner}, {McDermid}, {Sarzi}, \& {van de Ven}}]{Peletier_2007}
{Peletier}, R.~F., {Falc{\'o}n-Barroso}, J., {Bacon}, R., {et~al.} 2007,
  \mnras, 379, 445

\bibitem[{{Pfenniger} \& {Norman}(1990)}]{pfenniger_1990}
{Pfenniger}, D. \& {Norman}, C. 1990, \apj, 363, 391

\bibitem[{{Randich} {et~al.}(2013){Randich}, {Gilmore}, \& {Gaia-ESO
  Consortium}}]{randich2013}
{Randich}, S., {Gilmore}, G., \& {Gaia-ESO Consortium}. 2013, The Messenger,
  154, 47

\bibitem[{{Rangwala} {et~al.}(2009){Rangwala}, {Williams}, \&
  {Stanek}}]{rangwala_2009}
{Rangwala}, N., {Williams}, T.~B., \& {Stanek}, K.~Z. 2009, \apj, 691, 1387

\bibitem[{{Rattenbury} {et~al.}(2007){Rattenbury}, {Mao}, {Sumi}, \&
  {Smith}}]{rattenbury_barra}
{Rattenbury}, N.~J., {Mao}, S., {Sumi}, T., \& {Smith}, M.~C. 2007, \mnras,
  378, 1064

\bibitem[{{Recio-Blanco} {et~al.}(2006){Recio-Blanco}, {Bijaoui}, \& {de
  Laverny}}]{Matisse}
{Recio-Blanco}, A., {Bijaoui}, A., \& {de Laverny}, P. 2006, \mnras, 370, 141

\bibitem[{{Recio-Blanco} {et~al.}(2014){Recio-Blanco}, {de Laverny},
  {Kordopatis}, {Helmi}, {Hill}, {Gilmore}, {Wyse}, {Adibekyan}, {Randich},
  {Asplund}, {Feltzing}, {Jeffries}, {Micela}, {Vallenari}, {Alfaro}, {Allende
  Prieto}, {Bensby}, {Bragaglia}, {Flaccomio}, {Koposov}, {Korn}, {Lanzafame},
  {Pancino}, {Smiljanic}, {Jackson}, {Lewis}, {Magrini}, {Morbidelli},
  {Prisinzano}, {Sacco}, {Worley}, {Hourihane}, {Bergemann}, {Costado},
  {Heiter}, {Joffre}, {Lardo}, {Lind}, \& {Maiorca}}]{recio-blanco2014}
{Recio-Blanco}, A., {de Laverny}, P., {Kordopatis}, G., {et~al.} 2014, \aap,
  567, A5

\bibitem[{{Rich} {et~al.}(2007{\natexlab{a}}){Rich}, {Origlia}, \&
  {Valenti}}]{rich_no_gradiente}
{Rich}, R.~M., {Origlia}, L., \& {Valenti}, E. 2007{\natexlab{a}}, \apjl, 665,
  L119

\bibitem[{{Rich} {et~al.}(2012){Rich}, {Origlia}, \&
  {Valenti}}]{Rich_2012_no_gradiente}
{Rich}, R.~M., {Origlia}, L., \& {Valenti}, E. 2012, \apj, 746, 59

\bibitem[{{Rich} {et~al.}(2007{\natexlab{b}}){Rich}, {Reitzel}, {Howard}, \&
  {Zhao}}]{Rich_brava_I}
{Rich}, R.~M., {Reitzel}, D.~B., {Howard}, C.~D., \& {Zhao}, H.
  2007{\natexlab{b}}, \apjl, 658, L29

\bibitem[{{Saha} {et~al.}(2012){Saha}, {Martinez-Valpuesta}, \&
  {Gerhard}}]{saha_m_valpuesta}
{Saha}, K., {Martinez-Valpuesta}, I., \& {Gerhard}, O. 2012, \mnras, 421, 333

\bibitem[{{Saito} {et~al.}(2012){Saito}, {Hempel}, {Minniti}, {Lucas},
  {Rejkuba}, {Toledo}, {Gonzalez}, {Alonso-Garc{\'{\i}}a}, {Irwin},
  {Gonzalez-Solares}, {Hodgkin}, {Lewis}, {Cross}, {Ivanov}, {Kerins},
  {Emerson}, {Soto}, {Am{\^o}res}, {Gurovich}, {D{\'e}k{\'a}ny}, {Angeloni},
  {Beamin}, {Catelan}, {Padilla}, {Zoccali}, {Pietrukowicz}, {Moni Bidin},
  {Mauro}, {Geisler}, {Folkes}, {Sale}, {Borissova}, {Kurtev}, {Ahumada},
  {Alonso}, {Adamson}, {Arias}, {Bandyopadhyay}, {Barb{\'a}}, {Barbuy},
  {Baume}, {Bedin}, {Bellini}, {Benjamin}, {Bica}, {Bonatto}, {Bronfman},
  {Carraro}, {Chen{\`e}}, {Clari{\'a}}, {Clarke}, {Contreras}, {Corvill{\'o}n},
  {de Grijs}, {Dias}, {Drew}, {Fari{\~n}a}, {Feinstein},
  {Fern{\'a}ndez-Laj{\'u}s}, {Gamen}, {Gieren}, {Goldman},
  {Gonz{\'a}lez-Fern{\'a}ndez}, {Grand}, {Gunthardt}, {Hambly}, {Hanson},
  {He{\l}miniak}, {Hoare}, {Huckvale}, {Jord{\'a}n}, {Kinemuchi}, {Longmore},
  {L{\'o}pez-Corredoira}, {Maccarone}, {Majaess}, {Mart{\'{\i}}n}, {Masetti},
  {Mennickent}, {Mirabel}, {Monaco}, {Morelli}, {Motta}, {Palma}, {Parisi},
  {Parker}, {Pe{\~n}aloza}, {Pietrzy{\'n}ski}, {Pignata}, {Popescu}, {Read},
  {Rojas}, {Roman-Lopes}, {Ruiz}, {Saviane}, {Schreiber}, {Schr{\"o}der},
  {Sharma}, {Smith}, {Sodr{\'e}}, {Stead}, {Stephens}, {Tamura}, {Tappert},
  {Thompson}, {Valenti}, {Vanzi}, {Walton}, {Weidmann}, \&
  {Zijlstra}}]{vvv_dr1}
{Saito}, R.~K., {Hempel}, M., {Minniti}, D., {et~al.} 2012, \aap, 537, A107

\bibitem[{{Saito} {et~al.}(2011){Saito}, {Zoccali}, {McWilliam}, {Minniti},
  {Gonzalez}, \& {Hill}}]{saito_x_shape}
{Saito}, R.~K., {Zoccali}, M., {McWilliam}, A., {et~al.} 2011, \aj, 142, 76

\bibitem[{{Scannapieco} \& {Tissera}(2003)}]{scannapieco_2003}
{Scannapieco}, C. \& {Tissera}, P.~B. 2003, \mnras, 338, 880

\bibitem[{{Schlegel} {et~al.}(1998){Schlegel}, {Finkbeiner}, \&
  {Davis}}]{schlegelMapas}
{Schlegel}, D.~J., {Finkbeiner}, D.~P., \& {Davis}, M. 1998, \apj, 500, 525

\bibitem[{{Shen} {et~al.}(2010){Shen}, {Rich}, {Kormendy}, {Howard}, {De
  Propris}, \& {Kunder}}]{shen2010}
{Shen}, J., {Rich}, R.~M., {Kormendy}, J., {et~al.} 2010, \apjl, 720, L72

\bibitem[{{Sofue} {et~al.}(2009){Sofue}, {Honma}, \&
  {Omodaka}}]{sofue_masa_bulbo}
{Sofue}, Y., {Honma}, M., \& {Omodaka}, T. 2009, \pasj, 61, 227

\bibitem[{{Uttenthaler} {et~al.}(2012){Uttenthaler}, {Schultheis}, {Nataf},
  {Robin}, {Lebzelter}, \& {Chen}}]{uttenthaler_2012}
{Uttenthaler}, S., {Schultheis}, M., {Nataf}, D.~M., {et~al.} 2012, \aap, 546,
  A57

\bibitem[{{V{\'a}squez} {et~al.}(2013){V{\'a}squez}, {Zoccali}, {Hill},
  {Renzini}, {Gonz{\'a}lez}, {Gardner}, {Debattista}, {Robin}, {Rejkuba},
  {Baffico}, {Monelli}, {Motta}, \& {Minniti}}]{sergio_x_kine}
{V{\'a}squez}, S., {Zoccali}, M., {Hill}, V., {et~al.} 2013, \aap, 555, A91

\bibitem[{{Wegg} \& {Gerhard}(2013)}]{wegg_mapas}
{Wegg}, C. \& {Gerhard}, O. 2013, \mnras, 435, 1874

\bibitem[{{Zhao}(1996)}]{zhao_barra}
{Zhao}, H. 1996, \mnras, 283, 149

\bibitem[{{Zoccali} {et~al.}(2014){Zoccali}, {Gonzalez}, {Vasquez}, {Hill},
  {Rejkuba}, {Valenti}, {Renzini}, {Rojas-Arriagada}, {Martinez-Valpuesta},
  {Babusiaux}, {Brown}, {Minniti}, \& {McWilliam}}]{gibs_paper}
{Zoccali}, M., {Gonzalez}, O.~A., {Vasquez}, S., {et~al.} 2014, \aap, 562, A66

\bibitem[{{Zoccali} {et~al.}(2008){Zoccali}, {Hill}, {Lecureur}, {Barbuy},
  {Renzini}, {Minniti}, {G{\'o}mez}, \& {Ortolani}}]{zoccali2008}
{Zoccali}, M., {Hill}, V., {Lecureur}, A., {et~al.} 2008, \aap, 486, 177

\end{thebibliography}

\end{document}